\documentclass{aastex701}

\usepackage{amsmath}	

\begin{document}

\title{The orbital anisotropy profile of the \textit{Gaia}-Sausage/Enceladus accretion remnant}

\shorttitle{The anisotropy profile of GS/E}

\author[orcid=0000-0001-8472-6404]{James M. M. Lane}
\affiliation{David A. Dunlap Department of Astronomy and Astrophysics, University of Toronto, 50 St. George Street, Toronto ON, M5S 3H4, Canada}
\email[show]{jm.lane@alumni.utoronto.ca}  

\author[orcid=0000-0001-6855-442X,sname='North America']{Jo Bovy} 
\affiliation{David A. Dunlap Department of Astronomy and Astrophysics, University of Toronto, 50 St. George Street, Toronto ON, M5S 3H4, Canada}
\email{jo.bovy@utoronto.ca}

\begin{abstract}

The \textit{Gaia}-Sausage/Enceladus (GS/E) accretion remnant is one of the most important stellar populations in the Milky Way halo. Recent simulation-based work has suggested that the anisotropy profiles of remnants like GS/E decline towards the center and may be well-fit by an Osipkov-Merritt-type distribution function (DF). We study the anisotropy profile of GS/E using a chemically-selected sample of stars from APOGEE~DR17 and \textit{Gaia}. We find that the anisotropy profile of GS/E is high and constant with $\beta \sim 0.9$ beyond 8~kpc, dropping to $\beta \sim 0.4$ at 2~kpc. We fit a two-component Osipkov-Merritt anisotropy profile to the GS/E data, finding that a superposition of profiles with scale radii $r_{\mathrm{a}}=2$~kpc and 547~kpc, with a mixture fraction $k_\mathrm{om} = 0.88$ provide a better fit to the data than a constant anisotropy profile. Using this new model, we re-assess the density profile and mass of the GS/E remnant from a previous work that assumed a contant anisotropy, finding an increase in the derived mass from $1.5\times 10^{8}~\mathrm{M}_{\odot}$ to $2.29 ^{+0.95}_{-0.63}\times 10^{8}~\mathrm{M}_{\odot}$. In general, the superposition Osipkov-Merritt DF more satisfactorily matches the kinematics of the GS/E remnant than the constant anisotropy DF, and in the future will form a more reliable basis for modelling the remnant. Additionally, these new constraints on the kinematics of GS/E near the Galactic center are an important measurement for any future theoretical or simulation-based investigation into its nature and origin.

\end{abstract}



\section{Introduction}
\label{sec:Intro}

Over the last few years, the field of Milky Way galactic archaeology has progressed at an impressive pace due to the combination of astrometric data from the \textit{Gaia Space Telescope} and wide-field spectroscopic surveys. A key finding of this era is that the Milky Way stellar halo is comprised in large part of the remnants of past merger events (e.g., see \citealt{helmi20} and \citealt{deason24} for recent reviews). Chief among these populations is \textit{Gaia}-Sausage/Enceladus (GS/E), a collection of relatively metal-rich stars on highly radial orbits that dominates the stellar halo near the Sun \citep{belokurov18,haywood18,helmi18}. The interpretation of GS/E as the remnant of a major accretion event has become well-established over the last few years \citep[an excellent accounting is presented by][]{deason24}.

While the chemistry of GS/E has received detailed study \citep[e.g.,][]{helmi18,vincenzo19,mackereth19a,monty20,feuillet20,horta23a,Plevne25a}, of equally great interest are the structural and kinematic properties of the remnant. The density profile of GS/E has been studied recently by \citet{han22} and \citet{lane23}, who both find it to be well-approximated as a triaxial broken power law, with inner and outer indices of $\sim [1,4]$ and a break radius of $\sim 20$~kpc. This ties into prior results which demonstrate that the density profile of the Milky Way stellar halo in aggregate exhibits a notable break around the same radius \citep[e.g.][]{deason11,deason18,mackereth20}. 

The kinematics of the GS/E remnant have also been studied, most notably using Gaussian mixtures \citep{lancaster19,fattahi19,iorio21}. The conclusion of such work is mainly that GS/E exhibits high, constant anisotropy $\beta \sim 0.9$, where $\beta$ is defined using the spherical velocity dispersions $\sigma_{[r,\phi,\theta]}$ as 
\begin{equation}
    \label{eq:anisotropy}
    \beta = 1- \frac{\sigma^{2}_{\phi} + \sigma^{2}_{\theta}}{2\sigma^{2}_{r}}.
\end{equation}
In these studies, the remainder of the stellar halo (a mixture of the \textit{in-situ} stellar halo and other accretion remnants less dominant at the Solar radius) typically has a nearly isotropic $\beta \sim 0.3$. So when modelling the kinematic properties of GS/E \citep[e.g.,][]{lane23}, or when studying GS/E analogs in cosmological or tailored N-body simulations \citep[e.g.,][]{loebman18,fattahi19,mackereth19a,amarante22}, it is generally assumed that the remnant has a constant anisotropy of $\beta \sim 0.9$.

Recently, \citet{lane25} studied remnants of significant accretion events around Milky Way analogs in the IllustrisTNG simulations. They found that remnants with high average anisotropy ($\beta \gtrsim 0.7$) characteristically have low anisotropy in the inner galaxy ($r \lesssim 10$~kpc) that rises to a constant, high value at larger radii. Those authors found that Osipkov-Merritt distribution functions \citep{osipkov79,merritt85} provided a good fit to radially-anisotropy remnants (and that specifically mixtures of Osipkov-Merritt DFs provided the best fit). They posed the question: what is the anisotropy profile of GS/E, and does it decrease in the inner Galaxy like the remants in IllustrisTNG? Tentative evidence for this can be seen in the results of \citet{iorio21}, who study a large sample of RR Lyrae variables using Gaussian mixtures. They find that the anisotropy of their high-anisotropy component decreases towards the inner Galaxy (at about a few~kpc radius), but also that the fraction of the stellar halo comprised by the high-anisotropy component also decreases in the inner Galaxy. Other relevant studies either do not directly assess the anisotropy of GS/E as a function of radius \citep[e.g.,][]{belokurov18,fattahi19}, or are limited in their ability to probe the inner Galaxy due to the nature of the tracer \citep[e.g.,][]{lancaster19}.

One issue with studying samples of GS/E stars defined by their kinematics (i.e., typically a cut is done on eccentricity, angular momentum, radial action, or some combination thereof) is that if the anisotropy of the remnant does decrease in the inner Galaxy then such samples will not contain these stars. When studying GS/E using Gaussian mixtures a similar problem may arise: if the anisotropy of the high-anisotropy component of the model tends to decrease towards the inner Galaxy, it may become confused and degenerate with the low anisotropy component \citep[i.e.,][]{iorio21}, clouding inference.  

In this paper, we study the anisotropy profile of GS/E using a sample of stars defined by the abundance trends of the remnant, which have become well-characterized in recent years \citep{helmi18,vincenzo19,monty20,feuillet20,naidu20,horta23a}. Such a sample should not exhibit anisotropy that simply reflects the selection method, as a sample selected using kinematics undoubtedly would, and will therefore offer the least biased test of the anisotropy profile of GS/E in the inner Galaxy. Our aim is to determine whether or not the anisotropy profile of GS/E remains high towards the inner Galaxy, or whether it drops towards isotropy as found for simulated GS/E remnants by \citet{lane25}. In \S~\ref{sec:data}, we introduce the APOGEE and Gaia data used. In \S~\ref{sec:gse-anisotropy}, we define an abundance-based selection for GS/E and compute the anisotropy profile. In \S~\ref{sec:fitting-dfs}, we fit distribution functions to the GS/E anisotropy profile as a way to provide a simple, approximate full phase-space distribution for the remnant. Finally, in \S~\ref{sec:discussion} and \S~\ref{sec:conclusion}, we end with a discussion of our results and summarize our findings.

\section{Data}
\label{sec:data}

We use spectroscopic data from the seventeenth data release \citep[DR17;][]{apogee_dr17} of the  Apache Point Galactic Evolution Experiment \citep[APOGEE;][]{apogee}, a component of the fourth iteration of the Sloan Digital Sky Survey \citep[SDSS-IV;][]{sdss4}. APOGEE provides high-resolution ($R \sim 22,500$) \textit{H}-band spectra for about $7 \times 10^{5}$ stars across both Northern and Southern hemispheres, including numerous pointings near the Galactic center, which is ideal for this study. Abundances and stellar parameters are provided by the APOGEE Stellar Parameters and Abundances Pipeline \citep[ASPCAP;][]{apogee_aspcap}. We supplement the spectroscopic data with heliocentric distances, Galactocentric positions and velocities, and integrals of motion from the DR17 \texttt{astroNN} Value Added Catalogue (VAC)\footnote{\url{https://www.sdss4.org/dr17/data_access/value-added-catalogs/?vac_id=the-astronn-catalog-of-abundances,-distances,-and-ages-for-apogee-dr17-stars}} \citep{leung19a,leung19b}. The astrometric data used to produce this catalogue comes from Data Release 3 \citep[DR3;][]{gaiadr3} of the \textit{Gaia Space Telescope} \citep{gaia}. The spectrophotometric distances in this catalogue are derived using a Bayesian neural network trained on APOGEE spectra with accurate \textit{Gaia} parallaxes \citep{leung19b}. Integrals of motion in the catalogue are computed using the `St\"{a}ckel fudge' method of \citet{binney12}, implemented following the approach of \citet{mackereth18} in \texttt{galpy} \citep{bovy15}. The assumed potential is \texttt{MWPotential2014} presented by \citet{bovy15}.

The APOGEE \texttt{allstar} file is loaded and the matched \texttt{astroNN} VAC data added in using the \texttt{apogee} package\footnote{\url{https://github.com/jobovy/apogee/}} \citep{bovy16}. The data are cleaned according to the following procedure. First, duplicate observations and commissioning data are removed from the APOGEE catalogue using targeting bits and observation coordinates (see \texttt{rmdups} and \texttt{rmcommissioning} keywords in \texttt{apogee.tools.read.allStar}). We take only stars with $1 < \log g < 4$ to obtain a sample of giants for which chemical abundances and \texttt{astroNN}  distances are most reliable. We then remove data that do not have measured [Fe/H], [Al/Fe], or [Mg/Fe] abundances, or stars for which the uncertainties on these abundances is greater than 0.2 dex. We also remove stars with uncertainty on the mean heliocentric radial velocity greater than 50~km~s$^{-1}$. We remove stars with the following flags set: \texttt{TEFF\_BAD}, \texttt{LOGG\_BAD}, \texttt{STAR\_BAD} (bits 16, 17, and 23 respectively in \texttt{ASPCAPFLAG}). We also remove stars for which the following flags are set: \texttt{VERY\_BRIGHT\_NEIGHBOR}, \texttt{LOW\_SNR} \texttt{PERSIST\_HIGH}, \texttt{PERSIST\_JUMP\_POS}, \texttt{PERSIST\_JUMP\_NEG}, and \texttt{SUSPECT\_RV\_COMBINATION} (bits 3, 4, 5, 12, 13, 14 respectively in \texttt{STARFLAG}). We obtain RUWE from the \textit{Gaia} DR3 catalogue and remove stars with values greater than 1.4, which can indicate issues with the astrometric solution. We also require that stars have finite eccentricities in the domain $[0,1)$ from the \texttt{astroNN}  VAC. 

Finally, we impose two spatial cuts on the data pursuant to our science goals. First, we remove stars within 2~kpc distance of the Galactic center. This serves to remove any bulge stars from the sample. While the chemical selection for GS/E presented in the next section should preclude the inclusion of bulge stars in the sample, this physical cut acts as an additional safeguard. Second, we remove all stars lying within the projected tidal radius of any Milky Way globular cluster. We take the catalogue of Milky Way globular clusters and their structural properties produced by \citet{baumgardt18}. While it would be reasonable to also remove stars lying in close proximity to dwarf satellites of the Milky Way, the Galactocentric spatial extent of our sample is largely $< 30$~kpc (see Figure~\ref{fig:anisotropy}), meaning contamination from dwarf galaxies is negligible. Throughout our analysis we visually inspect the data to ensure no unreasonably large groups of stars with comparable chemistry or velocities are found, which gives us confidence that no unwanted substructure is present in the final catalogue.

\section{GS/E and its kinematics}
\label{sec:gse-anisotropy}

\subsection{GS/E selection}
\label{subsec:gse-selection}

With regards to choosing stellar samples, GS/E is commonly defined via its characteristic kinematics, which are very radially anisotropic. Energy, angular momentum, actions, and eccentricity are all used to create selection criteria for the remnant (see, e.g., \citealt{lane22} and \citealt{carrillo24} for comparisons of selection criteria). When selecting a GS/E sample on the basis of its kinematics, selections typically try to isolate halo stars on the most radial orbits. But such selections will obviously introduce kinematic bias into the resulting sample. For most scientific purposes, such as to study the chemistry or density profile of the remnant, this bias does not impact analysis of the resulting sample. But if the goal is to study the intrinsic kinematics of the remnant then care must be taken not to introduce bias by selecting only the most radially anisotropic constituents, which may miss any part of the remnant with less extreme kinematics.

With the goal of determining whether the anisotropy profile of GS/E drops in the inner Galaxy, we select GS/E  members using chemistry, which should result in a sample with less bias with regards to kinematics than a sample selected as outlined above. As is characteristic for the remnant of a dwarf galaxy, GS/E is elevated in [$\alpha$/Fe] at low [Fe/H], pivoting towards lower [$\alpha$/Fe] as [Fe/H] increases, the standard ``plateau-knee'' morphology \citep{helmi18,vincenzo19,monty20,hasselquist21,horta23a}. GS/E occupies a somewhat unique locus in [$\alpha$/Fe] versus [Fe/H] in the Milky Way, though, as it extends to much higher [Fe/H] \citep[about $-0.8$; e.g.,][]{hasselquist21} than many other constituent populations of the stellar halo. Moreover, at such high [Fe/H] it has low [$\alpha$/Fe] so that it separates nicely from \textit{in-situ} populations such as the metal-poor thick disk and the putative \textit{in-situ} stellar halo which have higher [$\alpha$/Fe] \citep[see][]{belokurov22,rix22}.

\begin{figure*}
    \centering
    \includegraphics[width=\textwidth]{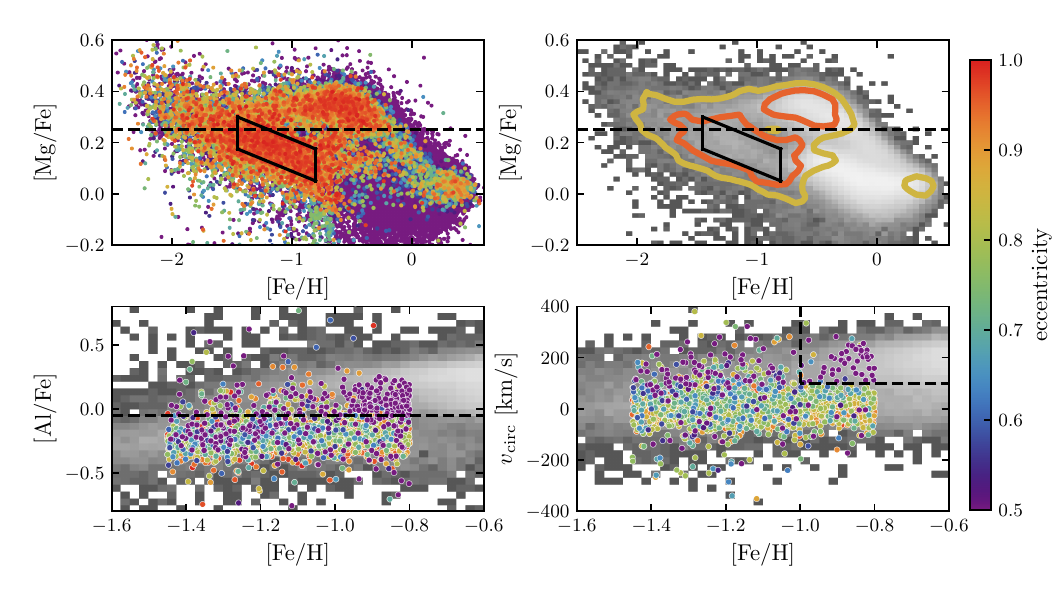}
    \caption{The selection of the GS/E sample. \textit{Top left:} [Mg/Fe] versus [Fe/H] for the entire cleaned APOGEE DR17 sample colored by eccentricity. The solid black box and dashed black line indicate the preliminary GS/E selections informed by eccentricity, see the text for details. \textit{Top right:} The Grayscale density of the entire APOGEE sample in [Fe/H] and [Mg/Fe]. The colored contours show the locations of loci of stars with $e>[0.85,0.95]$, the dashed black line is as in the first panel. \textit{Bottom left:} The preliminary chemical GS/E sample shown in [Al/Fe] versus [Fe/H] and colored by eccentricity, with the Grayscale density of the entire sample behind. The dashed black line indicates the threshold of $[\mathrm{Al/Fe}] < -0.05$ adopted to separate \textit{in-situ} and accreted stars. \textit{Bottom right:} The penultimate GS/E sample defined by the selections shown in the other panels of the figure, shown in Galactocentric circular velocity versus [Fe/H], colored by eccentricity, and with the Grayscale density of the entire smaple in the background. The dashed black line shows the the region occupied by likely disk star contaminants, which is excluded from the final sample.}
    \label{fig:apogee-feh-alpha}
\end{figure*}

We begin in the space of [Mg/Fe] versus [Fe/H] and, in order to identify the general chemical locus of GS/E, we examine the density of stars as a function of their eccentricity. The location of the highest eccentricity stars informs our first abundance-based selection for GS/E. This strategy rests on the assumption that GS/E stars with high eccentricity occupy the same locus in abundance space as GS/E stars with low eccentricity, an assumption we will discuss in more detail later. The top two panels of Figure~\ref{fig:apogee-feh-alpha} show the [Fe/H] and [Mg/Fe] abundances for the cleaned sample of APOGEE stars. Overlaid on the top right panel are contours showing the location of stars with eccentricity greater than 0.85 and 0.95 (colors are a drawn from the colorbar), which paint a clear picture of the location of GS/E in this space. The obvious location of interest here is the region of highest eccentricity at lower [Fe/H] and [Mg/Fe] \citep[c.f.,][]{hasselquist21,lane23,horta23a}. The collection of high-eccentricity stars at larger [Fe/H] and [Mg/Fe] is the hot part of high-$\alpha$ disk. The first GS/E sample criterion encompasses the high-eccentricity locus of interest using an area encompassed by 4 points, which we define as 
\begin{equation}
\mathcal{A} = \{ (\mathrm{[Fe/H]},\mathrm{[Mg/Fe]}) \lvert (-1.45,0.3), (-1.45,0.175), 
   (-0.8,0.05),  (-0.8,0.175)\}
\end{equation}
This area is shown in solid black in the top two panels of Figure~\ref{fig:apogee-feh-alpha}. In order to minimize contamination from \textit{in-situ} stars, which have higher $\alpha$ abundances comparable to the high-$\alpha$ disk \citep{das20,belokurov22}, we impose a ceiling at [Mg/Fe] = 0.25, shown as a dashed line in the top two panels of Figure~\ref{fig:apogee-feh-alpha}. The initial GS/E sample, which we refer to as $\mathcal{S}_{1}$, is the combination of these two criteria. 

Moving forwards, we check the degree of contamination from \textit{in-situ} stars by examining the [Al/Fe] abundances of the sample defined by $\mathcal{S}_{1}$; these [Al/Fe] abundances are shown in the bottom left panel of Figure~\ref{fig:apogee-feh-alpha}. We expect some contamination of our sample from non-halo stars based on the chemical selection $\mathcal{S}_{1}$ because the highest [Fe/H] tail of the GS/E distribution around $[\mathrm{Fe/H}]=-0.8$ overlaps with the low-[Fe/H] tail of the low-$\alpha$ disk. Al abundances are well-suited to differentiate accreted and \textit{in-situ} stars \citep{das20,belokurov22,ernandes25}. This is because Al yields are metallicity-dependent, making present-day Al abundances very sensitive to the mass of the system within which the stars were formed \citep[see a discussion on the use of Al in][]{belokurov22}. Indeed, examining Figure~\ref{fig:apogee-feh-alpha} we see that there exist many stars with low eccentricity at high [Al/Fe] concentrated towards the high-[Fe/H] end of the distribution. These are clearly low-$\alpha$ disk stars which contaminate the sample. In order to remove these stars, as well as any \textit{in-situ} stars likely present at lower [Fe/H] also with high [Al/Fe], we impose a cut at $[\mathrm{Al/Fe}]=-0.05$ \citep[see][for similar cuts]{belokurov22,lane23}.

The Galactocentric circular velocity of the stars selected using $\mathcal{S}_{1}$ and [Al/Fe] cuts are shown in the bottom right panel of Figure~\ref{fig:apogee-feh-alpha}. Here, we see a concentration of stars with high [Fe/H] and high $v_\mathrm{circ}$ (about 200~km~s$^{-1}$), which is clearly indicative of additional disk stars which have not been removed using the [Al/Fe] cut (see also \citealt{feuillet22}). As a final step we remove all stars lying within the box defined by $[\mathrm{Fe/H}]=-1$ and $v_\mathrm{circ}=100$~km~s$^{-1}$. Here, we are safe to employ kinematics as a selection mechanism since we do not see a population at comparable [Fe/H] with negative $v_\mathrm{circ}$ of the same magnitude, and therefore we can be confident that this is disk contamination and not a genuine part of the stellar halo. To confirm that these are certainly thin disk stars we also examine their $z_\mathrm{max}$ (the maximum excursion above and below the disk plane), finding that they have typical values of about $2-4$~kpc, as well as eccentricities less than 0.1. Thus by chemistry and kinematics they are clearly members of the disk.

We do not choose to remove a similar population of stars at lower [Fe/H] with low eccentricity and $v_\mathrm{circ} = 100-200\,\mathrm{km\,s}^{-1}$, even though there does not appear to be a comparable population at negative $v_\mathrm{circ}$. This is because these stars overlap a region of chemodynamical space that is well occupied by other halo stars, reducing confidence that they are genuine outliers as oppossed to perhaps a low-eccentricity tail of the GS/E population. Moreover, their eccentricities range from 0.2-0.4 and typical $z_\mathrm{max}$ values are about 5~kpc, so they do not exhibit such convincing disk kinematics as the group of high-[Fe/H] rotating stars. We are hesitant to attempt to use kinematic cuts with such precision when the nature of the stars being removed is not clear cut, and so we leave these stars in the sample. We discuss this choice, and investigate its impact on our results---which is minor---in \S~\ref{subsec:selection-discussion}.

Together, our selection can be summarized with the following expression, which we apply to the cleaned APOGEE DR17 data
\begin{equation}
        \big( (\mathrm{[Fe/H]},\mathrm{[Mg/Fe]}) \in \mathcal{A} \big) 
        \wedge~  \big( \mathrm{[Mg/Fe]} < 0.25 \big) \\
        \wedge~   \big( \mathrm{[Al/Fe]} < -0.05 \big) 
        \wedge~  \big( (v_\mathrm{circ} < 100\, \mathrm{km\,s}^{-1}) \lor (\mathrm{[Fe/H]} < -1) \big).
\end{equation}
\noindent The final sample of high-likelihood GS/E stars in the sample defined by this selection is 2,367.

\begin{figure}
    \centering
    \includegraphics[width=0.55\columnwidth]{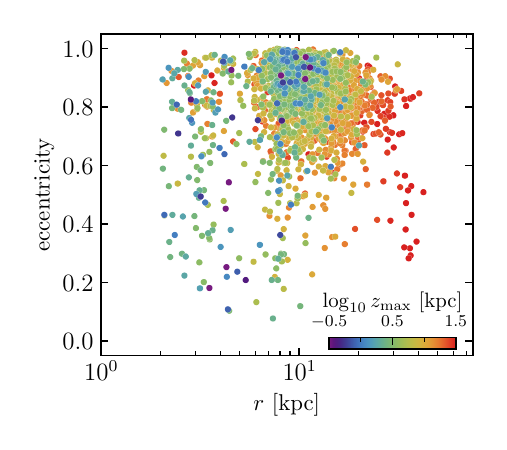}
    \caption{Eccentricity versus radius for the final GS/E sample used in our analysis, colored by $z_\mathrm{max}$.}
    \label{fig:gse-r-ecc-zmax}
\end{figure}

Figure~\ref{fig:gse-r-ecc-zmax} shows eccentricity as a function of radius for the final GS/E sample, colored by $z_\mathrm{max}$. This figure shows the final sample from a different perspective than Figure~\ref{fig:apogee-feh-alpha}. The distribution of eccentricities is clearly indicative of a halo population, with very few low-eccentricity stars, and the majority bearing a large $z_\mathrm{max}$. The combination of eccentricity and $z_\mathrm{max}$ is useful for distinguishing between halo and disk stars, because the latter would be expected to have both eccentricity and $z_\mathrm{max}$ near zero. 

To provide a proper gauge of the expected values of eccentricity and $z_\mathrm{max}$ for disk stars, we examine the distributions of each in the APOGEE DR17 sample. We find that the distribution of typical low-[Mg/Fe], high-[Fe/H] disk stars has eccentricities ranging from around 0 to 0.3 with a median value of about 0.15, and $z_\mathrm{max}$ values ranging from around 0.1 to 1~kpc with a median value of about 0.3~kpc. For typical high-[Mg/Fe], high-[Fe/H] disk stars, the eccentricities range from around 0 to 0.5 with a median value of about 0.25 and the $z_\mathrm{max}$ values range from around 0.3 to 2~kpc with a median value of about 1~kpc.

With this context in mind it becomes apparent that there are a handful of stars in the final GS/E sample with kinematics consistent with the Galactic disk. Specifically stars with eccentricity less than 0.5 and $\log_{10}\left(z_\mathrm{max}/\mathrm{kpc}\right)$ less than 0.5. But the total number of stars that meet both of these criteria is 39, out of a total sample of 2,367. So we can say that the possible disk contamination in the final sample is minimal. We will revisit Figure~\ref{fig:gse-r-ecc-zmax} later in the discussion when we assess contamination by disk stars in a more practical manner.

\subsection{The anisotropy profile of GS/E}
\label{subsec:anisotropy-profile-gse}

\begin{figure*}
    \centering
    \includegraphics[width=\textwidth]{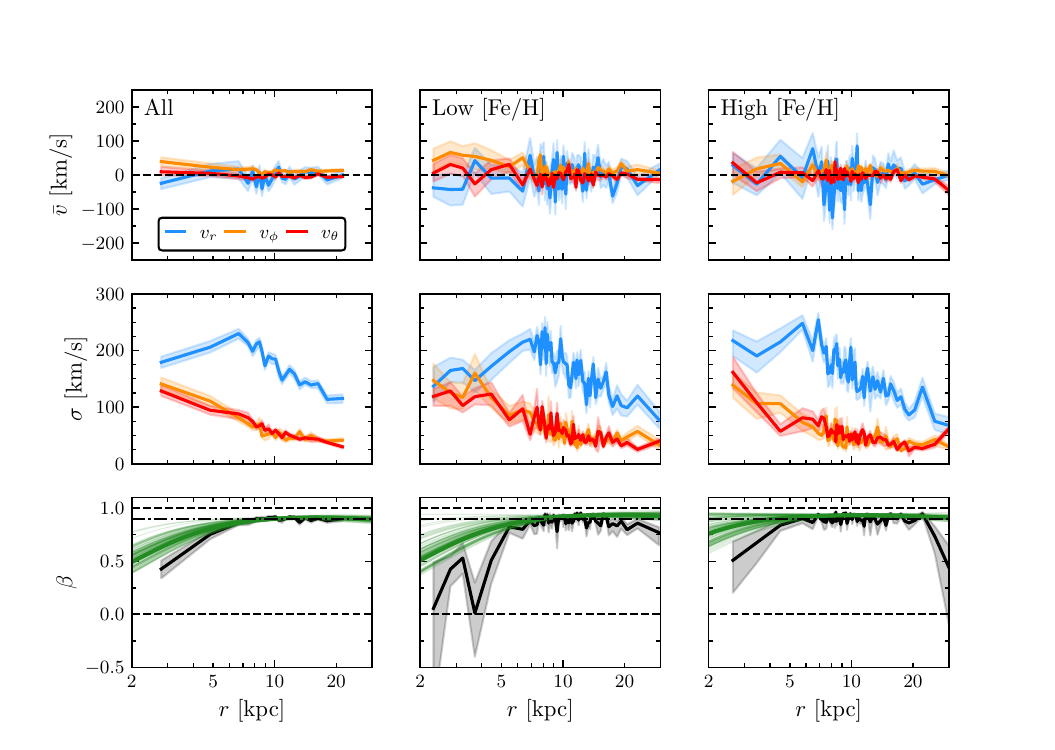}
    \caption{The kinematics of the radially-binned GS/E sample, as well as two subsamples defined by [Fe/H] abundance. The first row shows mean velocities in each spherical coordinate (colored), and the dashed line marks zero. The second row shows velocity dispersions using the same color scheme. The third row shows the anisotropy computed using Equation~\eqref{eq:anisotropy}. In these panels the two dashed lines mark $\beta=0$ and $\beta=1$, and the dash-dot line marks $\beta=0.9$, an important fiducial value for GS/E as it is the generally accepted value for its anisotropy at and beyond the Solar circle. For each quantity in each panel the solid line and fill shows the median and central 68th percentile of 100 bootstrap realizations of the trend. In the third row the green lines are samples from the posterior of the best-fitting mixture Osipkov-Merritt model (see \S~\ref{sec:fitting-dfs}). The second and third columns show subsets of the data when divided evenly in half by [Fe/H] abundance.}
    \label{fig:anisotropy}
\end{figure*}

We now assess the kinematics of our GS/E sample, which are shown in Figure~\ref{fig:anisotropy}. We begin by binning the data in radius, by definining bin edges such that there are 20 data points per bin. We then compute the mean velocity and velocity dispersion in each bin, and repeat the process by bootstrap resampling the original GS/E sample and recomputing velocity means and dispersions (the radial grid is fixed to the original throughout each bootstrap trial). For each set of velocity dispersions we compute the anisotropy following Equation~\eqref{eq:anisotropy}.

The resulting kinematics are shown in the first column of Figure~\ref{fig:anisotropy}. In the first panel, the mean velocities cluster nicely around zero at all radii. In the innermost bins there are fluctuations of a sufficient calibre to potentially indicate net streaming motions, however these are typically consistent with zero within the derived uncertainties. The velocity dispersions and anisotropy are shown in the second and third rows, and clearly indicate an Osipkov-Merritt-like profile. The anisotropy is high ($\beta \sim 0.9$) near the solar position and extending to the far radial limit of the sample, which is consistent with previous understanding of the GS/E remnants kinematics \citep{belokurov18,lancaster19,iorio21}. But in the inner Galaxy ($r \lesssim 8$~kpc) the anisotropy profile begins to drop, reaching $\beta=0.4$ around $2$ to $3$~kpc.

We also separate the GS/E sample into two equal-sized subsamples by [Fe/H]. To the resulting low- and high-[Fe/H] samples, we apply the same method to compute the mean velocity, velocity dispersion, and anisotropy profiles along with uncertainties. These are shown in the second and third columns of Figure~\ref{fig:anisotropy}. The purpose of dividing the sample by [Fe/H] is twofold. First to to assess the potential impact of any contamination on the GS/E sample, since for example contamination from the thin disk would only impact the high-[Fe/H] end of the GS/E sample, while the \textit{in-situ} halo may contaminate more prevalently at low-[Fe/H]. Second, there may be an intrinsic radial abundance gradient in the GS/E progenitor satellite which now persists as an abundance gradient of the remnant in the Milky Way. 

Indeed, there do appear to be some differences in the anisotropy profiles of the low- and high-[Fe/H] GS/E samples. The low-[Fe/H] sample has an anisotropy profile which approaches zero more smoothly in the inner Galaxy, beginning its decrease at approximately the same radius as does the profile for the whole GS/E sample ($\sim 8$~kpc). The high-[Fe/H] sample, on the other hand, only begins decreasing in anisotropy at the innermost radii probed by the sample, within 4~kpc. The anisotropy profiles of both subsamples track a high, constant value of $\beta=0.9$ at the solar circle and towards larger radii, consistent with the whole sample and expectations from prior studies of GS/E. The differences in the anisotropy profiles between the high- and low-[Fe/H] subsamples will be further discussed in \S~\ref{sec:discussion}.

\section{Fitting Osipkov-Merritt models to GS/E}
\label{sec:fitting-dfs}

\subsection{The procedure}

While the GS/E remnant's density is triaxial \citep[e.g.,][]{han22,lane23} and the Milky Way potential is aspherical, in this section we fit spherical Osipkov-Merritt models in a sphericalized version of the Milky Way's potential to the anisotropy profiles derived in the previous section. While it can only be approximate, this procedure allows us to provide a simple, yet useful full phase-space distribution for the GS/E remnant that get some of its main attributes correct (radial density profile and the anisotropy profile), even if it does not account for the non-spherical shape of the remnant. A justification for this procedure comes from \citet{lane25}, who found that realistic GS/E analogs in a set of Milky-Way-like galaxies in the IllustrisTNG simulation suite \citep{tng_public_release_nelson19} can we modeled well using equilibrium Osipkov-Merritt-type models in sphericalized versions of their host potential.

To fit spherical Osipkov-Merritt models, we follow the procedure described by \citet{lane25} and we provide a brief outline here. The interested reader may seek to consult Section~3 of that work for additional information.

Below, we fit two types of Osipkov-Merritt models: the original model first put forward independently by \citet{osipkov79} and \citet{merritt85} and a second model which consists of a superposition of two Osipkov-Merritt distribution functions; we refer to the original model as a ``single-component Osipkov-Merritt model`` and the second model as the ``superposition Osipkov-Merritt model''. While both are full DFs for the phase-space distribution, for the single-component model, the anisotropy profile is independent of the density profile or the gravitational potential and therefore stands on its own. For the superposition Osipkov-Merritt model, the anisotropy model requires the full DF and we therefore need to specify the host potential $\Phi$ and the tracer density profile $\rho$ to be able to calculate the anisotropy profile. Using established models of for the gravitational potential and GS/E tracer density, all of our Osipkov-Merritt-type models become full DF models, but we are here only interested in their anisotropies.

When we need to specify it, the adopted potential is \texttt{MWPotential2014} from \citet{bovy15}, consisting of three component potentials representing the stellar bulge, stellar disks, and dark halo. Because the models we fit are explicitly spherical, we construct a spherical representation of this potential by spline-interpolating the radial gravitational force $F_{r} = -GM(r)/r^{2}$ in the Galactic plane using the \texttt{interpSphericalPotential} class found in \texttt{galpy}. The tracer density profile of GS/E has been studied by \citet{han22} and \citet{lane23}, and a two power density profile with inner and outer power law slopes of 1 and 4 \citep[i.e., the model of][]{hernquist90} and a scale radius of 20~kpc is broadly consistent with both sets of results. We therefore assume this form for the density profile of GS/E when required.

The single-component Osipkov-Merritt model is similar to the ergodic DF given by the integral inversion derived by \citet{eddington16}, but with a modified relative energy 
\begin{equation}
    \mathcal{Q} = \mathcal{E} - \frac{L^{2}}{2r_{\mathrm{a}}}
\end{equation}
where $\mathcal{E} = -\Phi - \Phi(\infty) - \frac{1}{2} v^{2}$ is the relative energy, $v$ is the total velocity, $L$ is the total angular momentum, and $r_{\mathrm{a}}$ is a scale radius. The anisotropy of this model takes the form
\begin{equation}
    \label{eq:osipkov-merritt-anisotropy}
    \beta(r) = \frac{r^{2}}{r^{2} + r_{\mathrm{a}}^{2}}.
\end{equation}
such that $\beta = 0.5$ at $r = r_{\mathrm{a}}$ and $\beta$ tends towards 0 and 1 in the limit as $r$ tends towards 0 and $\infty$ respectively. The parameter $r_{\mathrm{a}}$ fully specifies the single-component Osipkov-Merritt anisotropy profile without reference to the potential or tracer density. 

\citet{merritt85} suggested a modified version of the Osipkov-Merritt model consisting of a superposition of two Osipkov-Merritt DFs with two different values of the scale radius $r_{\mathrm{a},1}$ and $r_{\mathrm{a},2}$. In this case the radial and tangential velocity dispersion for the combined model would be 
\begin{equation}
    \label{eq:superposition-osipkov-merritt-dispersions}
    \begin{split}
        \sigma_{r}^{2} = &\ k_\mathrm{om} \sigma_{r,1}^{2} + (1-k_{\mathrm{om}}) \sigma_{r,2}^{2} \\      
        \sigma_{t}^{2} = &\ k_\mathrm{om} \sigma_{t,1}^{2} + (1-k_{\mathrm{om}}) \sigma_{t,2}^{2}\,, \\      
    \end{split}
\end{equation}
where $k_\mathrm{om}$ is the relative fraction of the Osipkov-Merritt models corresponding to $r_{\mathrm{a},1}$ and $r_{\mathrm{a},2}$. Because computing the anisotropy profile for this model requires fully computing the radial and tangential velocity dispersions, the anisotropy profile depends on the potential and tracer density in addition to each components $r_{\mathrm{a}}$. 

\citet{lane25} studied both types of Osipkov-Merritt models for the purpose of serving as DFs for merger remnants in the halos of Milky Way analogs found in the IllustrisTNG simulation suite \citep{tng_public_release_nelson19,tng50_nelson19,tng50_pillepich19}. They found that for remnants with high average anisotropy (e.g., GS/E) Osipkov-Merritt DFs outperformed alternatives, and in particular the superposition DF described here was found to be superior to the standard single component Osipkov-Merritt DF. Here, we will fit both of these models to the data, including the single component Osipkov-Merritt profile due to its simplicity.

When fitting both models, we sample from a posterior constructed using Bayes rule $p(\theta \lvert \mathcal{D}) \propto p(\mathcal{D} \lvert \theta)p(\theta)$, where $p(\mathcal{D} \lvert \theta)$ is the model likelihood and $p(\theta)$ are any priors. In general, we use a simple Gaussian objective function to act as the likelihood between model anisotropy and data. Relevant priors will be described for each model below. Samples are drawn using the Markov Chain Monte Carlo (MCMC) sampler of \citet{goodman10} implemented in the package \texttt{emcee} \citep{foreman-mackey13}. When fitting models, we first minimize the log-likelihood function using Powell's conjugate direction method to determine the initial conditions for the sampler. We initialize 100 walkers in a small volume about this minimum.

To fit the single component Osipkov-Merritt profile to data is simple. All that needs to be determined is $r_{\mathrm{a}}$. To do this, we take the anisotropy profile for GS/E, computed above, and fit the anisotropy profile specified by equation~\eqref{eq:osipkov-merritt-anisotropy}. The likelihood $p(\mathcal{D} \lvert \theta)$ here is simple the Gaussian difference between the model and data. Adding the assumed gravitational potential and tracer density makes this a full DF model for GS/E, but we stress that we do not fit for the potential and/or tracer density here.

The procedure to fit the superposition Osipkov-Merritt model is more involved, since the anisotropy is indirectly specified by the velocity dispersions for each model, as per equation~\eqref{eq:superposition-osipkov-merritt-dispersions} and therefore requires the potential and tracer density and to perform the DF inversion. For even simple models such as Osipkov-Merritt DFs, it is computationally inconvenient to, on the fly, compute the DF and the relevant velocity dispersions for the anisotropy profile. We navigate this issue by computing $\sigma_{r}$ and $\sigma_{t}$ on a grid of $r$ and $r_{\mathrm{a}}$ before fitting. We consider 100 values of $r_{\mathrm{a}}$ logarithmically spaced from $10^{-1}$~kpc to $10^{2.5}$~kpc and 20 values of $r$ logarithmically spaced from 1 to 50~kpc (encompassing the radial range of the data). Using interpolation we can then quickly compute combined model velocity dispersions using equation~\eqref{eq:superposition-osipkov-merritt-dispersions} for any $r_{\mathrm{a}}$ at any radius while fitting. We examine the velocity dispersion grids by-eye, and find them to appear well-behaved and smooth, such that we are comfortable using interpolation in this situation. To fit we follow the same approach as used for the single component model above, employing a model likelihood consisting of the Gaussian difference between model and data. 

\subsection{Fitting results}

We present the results of our fits in Table~\ref{tab:om-fitting-results}. We also perform this same analysis on the sample of stars split by [Fe/H], which is described above, for both the normal Osipkov-Merritt and superposition Osipkov-Merritt model. The superposition Osipkov-Merritt results show reasonable uniformity between different abundance-based samples (All, Low [Fe/H], High [Fe/H]), whereby the inner $r_{\mathrm{a},1}$ are all within $\pm 1$~kpc, the outer $r_{\mathrm{a},2}$ are all within about 5~per~cent of one another, and the values of $k_\mathrm{om}$ are within 0.03 of one another. This helps to solidify the notion that the GS/E subsample behaves as a unified stellar population, since dividing it into abundance-defined subsamples was meant to help reveal whether contamination may impact the sample. The superposition Osipkov-Merritt model is compared to the data for the full and [Fe/H] split samples in Figure~\ref{fig:anisotropy}.

The standard Osipkov-Merritt results  for the different [Fe/H] cuts are also similar to one another, and similar to the superposition Osipkov-Merritt results. These $r_{\mathrm{a}}$ are systematically higher, which is to be expected because the single Osipkov-Merritt anisotropy profile asymptotically approaches $\beta=1$ at large radii. Therefore, a slightly larger value of $r_{\mathrm{a}}$ is needed to help fit the data which plateaus at $\beta=0.9$. Interestingly, since the superposition Osipkov-Merritt anisotropy profile is unable to decrease fast enough at small radii (see Figure~\ref{fig:anisotropy}), the worst part of the fit is at small radii. The standard Osipkov-Merritt anisotropy profile actually matches the data quite well in the inner Galaxy, but since it asymptotes at $\beta=1$ it is a poor match beyond 8~kpc, where the majority of the data is located.

Figure~\ref{fig:anisotropy} demonstrates that the superposition Osipkov-Merritt model is able to capture the $\beta \approx 0.9$ large-radius behavior well, but does not match the steep decline towards at $r < 5\,\mathrm{kpc}$ towards $r = 0$. As discussed in Section~\ref{subsubsec:selection-choices-discussion} below, a small part of this may be due to disk contamination, but the effect is too small to fully explain the discrepancy between the model and the data. Thus, it is likely that the DF of GS/E approaches isotropy faster than the superposition Osipkov-Merritt model employed here and it would be interesting to fully empirically determine the GS/E DF.

\begin{deluxetable}{llll}
\tablecolumns{4}
\tablehead{\colhead{Sample} & \colhead{$r_{\mathrm{a},1}$~[kpc]} & \colhead{$r_{\mathrm{a},2}$~[kpc]} & \colhead{$k_\mathrm{om}$}}
\tablecaption{Best-fitting parameters for the superposition Osipkov-Merrit model.\label{tab:om-fitting-results}}
\startdata
\cutinhead{Superposition Osipkov-Merritt model} \\
All & $1.99^{2.30}_{1.64}$ & $547.21^{851.09}_{212.34}$ & $0.88^{0.90}_{0.86}$ \\ 
Low [Fe/H] & $1.94^{2.31}_{1.48}$ & $526.27^{846.86}_{194.98}$ & $0.92^{0.94}_{0.89}$ \\ 
High [Fe/H] & $0.92^{1.35}_{0.40}$ & $516.29^{844.97}_{184.62}$ & $0.89^{0.91}_{0.87}$ \\ 
\cutinhead{Osipkov-Merritt model} \\
All & $3.20^{3.33}_{3.07}$ & - & - \\ 
Low [Fe/H] & $2.87^{3.02}_{2.71}$ & - & - \\ 
High [Fe/H] & $2.34^{2.46}_{2.22}$ & - & - \\     
\enddata
\end{deluxetable}

\section{Discussion}
\label{sec:discussion}

\subsection{The GS/E selection}
\label{subsec:selection-discussion}

One of the key steps in this analysis is the preparation of the GS/E sample with which we compute and fit the anisotropy profile. The main result, that the orbital anisotropy of GS/E falls at small radii and that it is, thus, better described by an Osipkov-Merritt-type model than a constant anisotropy model, obviously depends on the purity of the GS/E sample in the inner Galaxy, where these two models differ the most. It is therefore worthwhile to consider how pure the sample may be, and whether or not contamination from another source with nominally lower anisotropy, such as the \textit{in-situ} halo, bulge, or thick disk, could cause issues.

In \S~\ref{subsec:gse-selection}, we lay out the selection procedure for GS/E, which focuses on chemistry to avoid introducing kinematic biases into the resulting sample. We begin with a selection in [Fe/H] and [Mg/Fe] which is motivated by previous literature results as well as the occurrence of high-eccentricity stars. Then [Al/Fe] is used to separate \textit{in-situ} and accreted stars. Finally, a number of stars with low eccentricity, high [Fe/H], and high positive (in the direction of Galactic rotation) $v_\mathrm{circ}$ are removed as obvious disk interlopers.

We present both a combined anisotropy profile as well as two profiles for sub-populations of the GS/E sample split by [Fe/H]. All samples demonstrate the characteristics of an Osipkov-Merritt model, whereby the anisotropy is elevated and approximately constant at large radii, and it decreases sharply at small radii. The nature of the anisotropy profiles is somewhat different when comparing the low and high-[Fe/H] subsamples. In the low-[Fe/H] subsample the anisotropy begins to drop at a slightly larger radius, giving it a slightly larger $r_{\mathrm{a}} \sim 2$~kpc as opposed to the $r_{\mathrm{a}} \sim 1$~kpc of the high-[Fe/H] sample. 

\subsubsection{Sources of the contamination in the GS/E sample}
\label{subsubsec:contamination-discussion}

Now it is useful to consider these results in the context of various possible sources of contamination. First and most obvious is the \textit{in-situ} stellar halo, consisting most prominently of the Aurora population in the inner Galaxy \citep{belokurov22,rix22}. This old halo population has been linked to the early formation of the Galaxy; it has nominally isotropic kinematics with net rotation beginning as [Fe/H] increases within the population. A key difference between Aurora and GS/E, however, is [Al/Fe] abundances, as Aurora is comparatively enriched in Al owing to its origin within the Milky Way \citep{belokurov22}. GS/E, in contrast, is less enriched in Al because its progenitor was most likely a dwarf galaxy in proximity to the larger Milky Way. This pattern arises because Al abundance enrichment is very sensitive to the size of the host galaxy in which the stars are formed. This makes Al superior at separating \textit{in-situ} from accreted populations when compared with other abundances whereby the enrichment trends with host galaxy size, such as Fe. Therefore it would be expected that the [Al/Fe] cuts employed in our GS/E selection protocol would be effective at removing contamination from the \textit{in-situ} halo in general, and Aurora in particular. In the next section, we will explore a scenario where the observed anisotropy profile is caused by a constant anisotropy GS/E plus contamination from a low anisotropy halo component like \textit{Aurora}.

Next, we consider the Galactic bulge as a possible source of contamination. This can be safely discounted owing to two steps in our GS/E selection protocol: the choice of locus in [Fe/H]-[Mg/Fe] space and the Galactocentric radial range over which the study is conducted. The bulge is enhanced in $\alpha$ elements and [Fe/H] \citep[for a thorough review of bulge chemistry see ][]{barbuy18}, which places it at the nearly opposite location in the [Mg/Fe]-[Fe/H] plane when compared with GS/E, which lies at low [Mg/Fe] and (comparatively) low [Fe/H]. In order to test whether or not bulge stars may nontheless leak into the sample, we repeat our data selection procedure but include all stars regardless of radial location. We find that only 37 (an increase of 1.6 per cent to the size of the sample) additional stars are added to the sample, indicating that even if we include the physical location of the bulge in our sample the contamination is still minimal.

Finally, we can consider the disk as a potential source of contamination. Both the high- and low-$\alpha$ disks could contaminate our sample at opposite ends of the GS/E chemical selection locus. The low-$\alpha$ disk, at its most metal poor extent, overlaps to a small extent at the high-[Fe/H] end of the GS/E distribution, while the high-$\alpha$ disk, at its most metal poor extent, overlaps at the low-[Fe/H] end of the GS/E distribution. Two steps serve to prevent contamination from these stars in our sample. First, we remove stars with high-[Al/Fe], which should exclude most of the low-$\alpha$ disk, which is enhanced in [Al/Fe] \citep[see e.g.,][]{belokurov22}. Second, we cut high-[Fe/H], high-$v_\mathrm{circ}$ stars which are obvious thin disk contaminants. These steps are certainly more focused on removing low-$\alpha$ disk stars, and examination of Figure~\ref{fig:apogee-feh-alpha} does indeed show an excess of low-[Fe/H] stars with $v_\mathrm{circ} \sim  100-200~\mathrm{km\,s}^{-1}$ and low eccentricity. The specific kinematics of these stars would suggest membership of the thick disk, but it is also possible they could be a part of other weakly rotating halo populations such as the \textit{Splash} \citep{belokurov20}. We do not attempt to further refine our selection to pick and choose among these stars as we are attempting to avoid explicit kinematic selections so as to gauge the anisotropy of GS/E as faithfully as possible (the low-eccentricity, high-[Fe/H], high-$v_\mathrm{circ}$ selection being an exception as those stars are an obvious group of disk stars with no other stars of apparent halo origin occupying the same region of phase space). In the next section, we will briefly examine the effect of our sample selection protocol on the resulting anisotropy profile, which will include exploring additional cuts which remove many of these stars.

\subsubsection{The impact of selection choices on the observed anisotropy profile}
\label{subsubsec:selection-choices-discussion}

We have argued here that the GS/E selection employed in this work is of a high fidelity, and that the stars which make up the sample used to assess the anisotropy profile are genuine members of GS/E. We can investigate the impact of the selection on our results by measuring the anisotropy profile using different subsets of stars. We begin with selections more broad than the one we present in \S~\ref{subsec:gse-selection} and then afterwards explore selections which are more restrictive. 

First, we examine the anisotropy profile, produced in the same manner as is done in \S~\ref{sec:gse-anisotropy}, for a sample of stars generated only using the [Fe/H]-[Mg/Fe] cuts (top left panel of Figure~\ref{fig:apogee-feh-alpha}). The resulting anisotropy profile has $\beta$ values about 0.1 lower systematically along its whole extent, except beyond 15~kpc, where the anisotropy drops to the range of 0.5-0.7. Examining the kinematics of the stars suggests that at intermediate radii where the curves diverge there is an excess of stars with $v_\mathrm{circ} \sim 200~\mathrm{km\,s}^{-1}$, $e \sim 0.1$, and high [Fe/H], suggesting disk star contamination. The overall decrease in the anisotropy across the curve suggests that the [Al/Fe] cuts, which are not included for this sample, are removing stars with more isotropic kinematics, consistent with the idea that these are \textit{in-situ} stars.

Next, we investigate the sample of stars produced using the [Fe/H]-[Mg/Fe] cuts as well as the [Al/Fe] cut (i.e. identical to our sample except for the $v_\mathrm{circ}$-[Fe/H] cut to remove the excess thin disk stars). Unsurprisingly since this sample is so similar to the final GS/E sample---they only differ by 56 stars---the resulting anisotropy profile is very similar to the profile shown in Figure~\ref{fig:anisotropy}. The anisotropy is lower by less than 0.05 beyond 5~kpc, and about 0.1 lower beyond 20~kpc. Building on the interpretation of the [Fe/H]-[Mg/Fe]-only sample discussed above, the behaviour of the anisotropy profile here is also consistent with contamination by thin disk stars, which the $v_\mathrm{circ}$-[Fe/H] cut removes.

We finally test cuts that are more restrictive than the one used in this study. We create samples identical to the one presented in \S~\ref{subsec:gse-selection} but with additional cuts on eccentricity greater than 0.2, 0.3, 0.4, and 0.5. The trend as the eccentricity cuts become more strict is for the resulting anisotropy profile to take higher values, especially within the inner 10~kpc. But the change is not as drastic as might be expected when compared with the profile in Figure~\ref{fig:anisotropy}. The $e > 0.2$ cut produces a sample with an anisotropy profile almost identical to that standard profile, and for the $e > 0.5$ sample the anisotropy only rises by about 0.05 at small radii, and is only marginally higher in value at large radii. Comparing these findings with the eccentricity distribution for the nominal GS/E sample shown in Figure~\ref{fig:gse-r-ecc-zmax} is illuminating. In that Figure it we can see that the low-eccentricity stars removed using these cuts (up to $e=0.5$) lie at smaller radii. We can also see that when only examining the stars with $e > 0.5$ there is still the rough shape of an Osipkov-Merritt distribution. At small radii there is a roughly even distribution of low- and high-eccentricity stars, and at larger radii the distribution is dominated by higher eccentricity stars.

Indeed, the fact that the anisotropy is slightly higher in the inner Galaxy when more aggressive eccentricity cuts are used may be positive, as the best-fitting superposition Osipkov-Merritt anisotropy profile is too high in the inner Galaxy (see Figure~\ref{fig:anisotropy}). This change is certainly consistent with the discussion above regarding potential contamination from low-eccentricity, high-$v_\mathrm{circ}$ thick disk stars in the final sample. But the fact that this change, as well as the others discussed here, produces very little change in the nature of the observed anisotropy profile should serve to demonstrate that our findings are robust to specific choices in the analysis. Regardless, we still believe it is important to study the anisotropy profile of GS/E without imposing kinematic cuts on the sample.

\subsection{Assessing the scenario of a constant anisotropy GS/E plus \textit{in-situ} contamination}

One possibility that could explain the observed anisotropy profile of GS/E presented here is that of a constant anisotropy GS/E plus low anisotropy, non-GS/E contamination. In this scenario, GS/E has a high, constant anisotropy all the way into the inner Galaxy, but the samples that we work with are contaminated by low anisotropy stars, likely from the \textit{in-situ} halo. While above we discuss how the chemical selections we employ likely remove much of that contamination, let us here entertain this hypothesis. We propose, and will here demonstrate, that while a scenario of constant anisotropy GS/E plus contamination may yield an anisotropy profile consistent with our data, more detailed inspection of the kinematics can distinguish between these two cases. To perform this analysis, we will draw samples from two sets of DFs representing this hypothesis and the best-fitting result from this work, and then study the distributions of their kinematics. 

We begin with two sets of DFs. The first is the best-fitting superposition Osipkov-Merritt DF from \S~\ref{sec:fitting-dfs}, which we refer to from now on as the OM2 model. The other DF is a set representing the alternate hypothesis we investigate here. This second set of DFs will be a superposition of two constant anisotropy DFs, one with $\beta=0.3$ representing the contamination and another with $\beta=0.9$ representing GS/E. In the contamination scenario, the GS/E stellar population has a constant anisotropy of $\beta=0.9$, which is well-supported by the literature \citep{belokurov18,lancaster19,iorio21}. If there was a centrally-concentrated stellar population with more ergodic, $\beta=0.3$ kinematics that contaminated the sample used in this study, then the resulting anisotropy profile may follow what is observed: $\beta$ trending towards 0 in the inner Galaxy and high and constant in the outer Galaxy. The \textit{in-situ} \textit{Aurora} population first studied by \citet{belokurov22} \citep[see also][]{rix22} may satisfy this requirement, as this population is centrally concentrated in the Milky Way and has more isotropic kinematics compared with GS/E. Moving forwards we will assume that the contaminant is \textit{Aurora}, even though we have no proof that this population is necessarily the contaminant in question for this hypothesis.

For the GS/E population, we use the same density profile assumed for the DF in \S~\ref{sec:fitting-dfs}, a Hernquist profile roughly matched to the findings of \citet{lane23}. For the Aurora population, we assume a power law density profile with index $\alpha=4.5$ \citep[consistent with the results of ][]{kurbatov24} and an exponential cutoff radius at 50~kpc \citep[not part of the density profile fit by ][but added for computational stability of the DF]{kurbatov24}. The DF has a constant anisotropy of $\beta=0.3$, consistent with estimates of the anisotropy for the \textit{in-situ} Milky Way stellar halo. From this point forward, we refer to this DF model as CA+Aur for convenience. Determining the relative normalization of the Aurora and GS/E DFs in the CA+Aur model is an obviously challenging question, because it strongly depends on the respective metallicity, spatial, and stellar type distributions of the GS/E and Aurora populations. We can roughly determine the mixture necessary in a data driven way such that the observed anisotropy profile shown in Figure~\ref{fig:anisotropy} is produced. We do this manually, by drawing samples from the constant anisotropy GS/E and Aurora DFs and varying their respective fractions until we are able to reproduce the observed anisotropy profile as closely as possible. The DF samples are equal in size to the real data (2,367 stars). The radial distribution of each DF sample also matches the radial distribution of the data, but since the DFs are spherical, distribution of the samples in azimuth or height above the Galactic plane are ignored. 

\begin{figure}
    \centering
    \includegraphics[width=0.55\columnwidth]{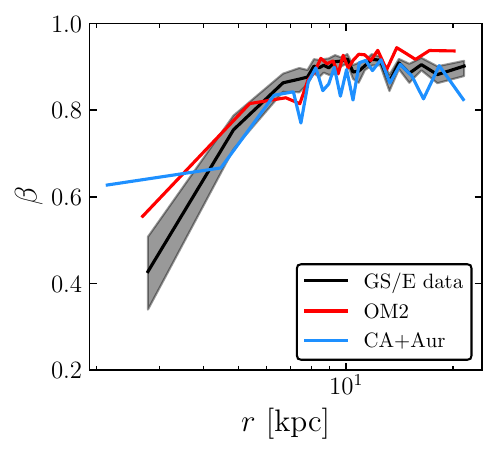}
    \caption{The anisotropy profiles of the mock samples and real data. The black curve shows the anisotropy profile of the APOGEE DR17 GS/E sample, which is the same as is shown in Figure~\ref{fig:anisotropy}, and the Gray fill around the curve represents the bootstrapped uncertainty. The red and blue curves are the anisotropy profiles of the OM2 and CA+Aur DF samples.}
    \label{fig:mock-anisotropy}
\end{figure}

We show the anisotropy profiles of the mock samples alongside the real data in Figure~\ref{fig:mock-anisotropy}. The red curve, representing the OM2 model, is clearly a good match to the data insofar as it corresponds to the posterior shown in Figure~\ref{fig:anisotropy}. The blue curve, representing the CA+Aur model, is a reasonable approximation to the data, and is at least close to the best-fitting anisotropy profile (i.e., it does not drop nearly fast enough as radius approaches zero). The anisotropy profiles of the two DF samples are reasonable matches to one another, indicating that the relative contribution of the two DFs in the CA+Aur model are appropriately normalized for this exercise.

\begin{figure*}
    \centering
    \includegraphics[width=\textwidth]{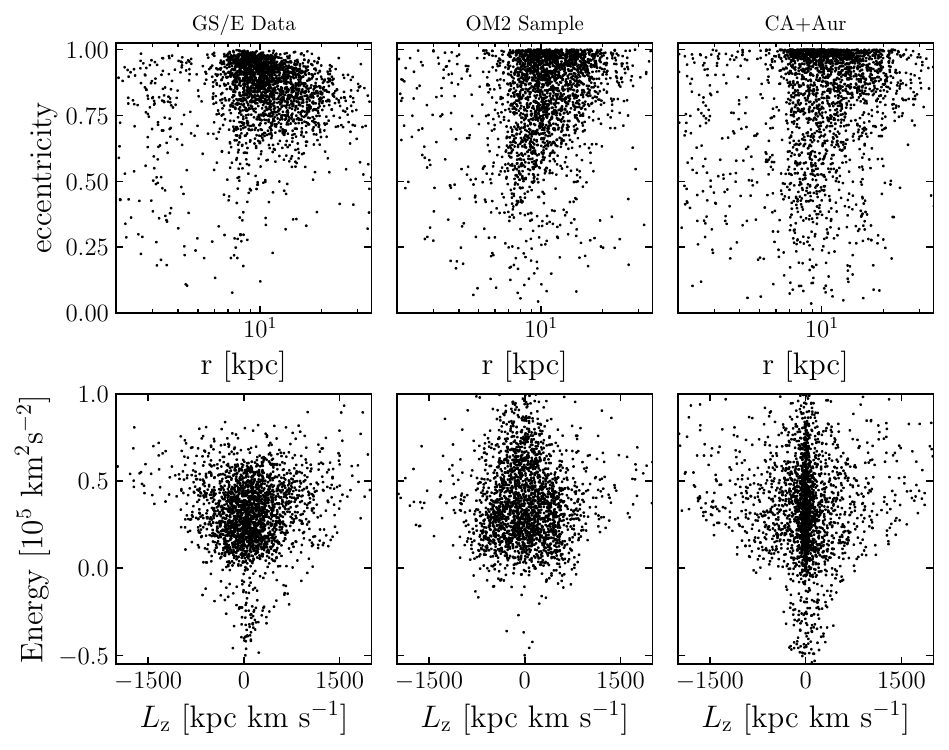}
    \caption{Kinematics of observed GS/E data (left column), OM2 DF sample (middle column), and CA+Aur DF (right column). \textit{Top row:} eccentricity versus Galactocentric radius, \textit{bottom row:} energy versus z-axis angular momentum.}
    \label{fig:mock-erELz}
\end{figure*}

We now turn to a more detailed investigation of the kinematics of the GS/E data and DF samples. We compute integrals of motion (eccentricity and actions) for the DF samples using the `St\"{a}ckel fudge' in the same manner as for the APOGEE DR17 data, as outlined in \S~\ref{sec:data}. Figure~\ref{fig:mock-erELz} shows some of these kinematics for the GS/E data as well as the DF samples. The top row shows eccentricity versus Galactocentric radius and the bottom row shows energy versus angular momentum. This figure begins to reveal to us that, while the anisotropy profiles of the two sets of DF samples are similar, they are more different when viewed in terms of other kinematic properties. The eccentricity distribution of the CA+Aur model is much more concentrated towards $e=1$ than the data shows. Similarly, this DF sample has a cuspy distribution in $L_\mathrm{z}$ that clearly does not match the data. The OM2 DF sample is more similar to the real data, as it appears to match the density of the eccentricity-radius and energy-angular momentum distributions better.

\begin{figure*}
    \centering
    \includegraphics[width=\textwidth]{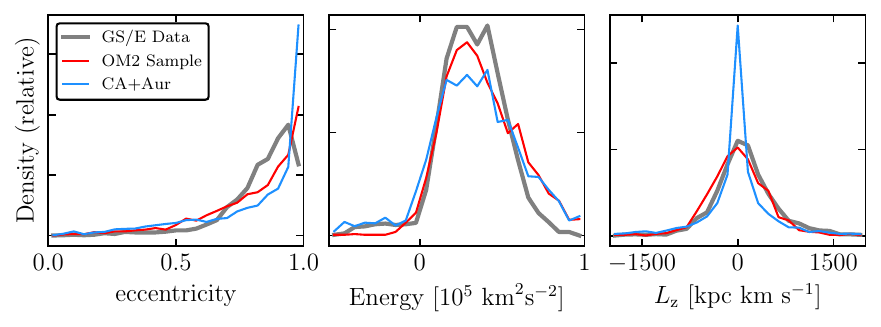}
    \caption{Marginal distributions of eccentricity, energy, and angular momentum for the GS/E data and DF samples, the same data shown in Figure~\ref{fig:mock-erELz}. The thick Gray line shows the GS/E data, and the red and blue lines show the combined Osipkov-Merritt and constant anisotropy GS/E plus Aurora DF samples respectively.}
    \label{fig:mock-erELz-margins}
\end{figure*}

Figure~\ref{fig:mock-erELz-margins} shows the marginal distributions of the quantities shown in Figure~\ref{fig:mock-erELz}: eccentricity, energy, and angular momentum. Examining first the energy distribution shows that, while both distributions are well matched to one another, neither is matched perfectly to the data. But the disparity is not large, and can likely be partially explained by the fact that the DFs use a spherical potential. The other two sets of kinematic parameters reveal more interesting differences between the two DF samples and their relation to the GS/E data. Beginning with eccentricity, both DF samples exhibit more cuspy behaviour near $e=1$ compared with the GS/E data, but the CA+Aur DF sample is more strongly cusped, and overall less well-matched to the GS/E data. The OM2 sample has roughly the correct amount of high-eccentricity stars, the distribution just peaks at $e=1$ rather than at a slightly lower value like the GS/E data. The angular momentum distribution is more definitive at showing the manner in which the CA+Aur DF sample does not provide an adequate match to the data. The GS/E data and the OM2 sample have gently peaked, approximately Gaussian distributions in angular momentum. The CA+Aur distribution is far more aggressively peaked, with over a factor of 2 more stars with $L_\mathrm{z} \approx 0$ compared with the GS/E data and OM2 sample. 

So we have shown that while both the OM2 and CA+Aur DFs may produce similar looking anisotropy profiles, upon deeper inspection of additional kinematic parameters it is revealed that the two DFs behave differently. We also saw that the OM2 DF appeared to be a better match to the GS/E data, especially when examining the marginal distributions of kinematic parameters, compared with the CA+Aur DF sample. The fact that the OM2 DF sample kinematics do not perfectly match those of the real data is not particularly surprising, and is the same finding as \citet{lane25} who studied simulated GS/E-like accretion remnants in a similar manner. So to conclude this exercise, it is clear that the OM2 DF better describes the kinematic properties of the GS/E data than does the CA+Aur DF. When combined with arguments laid out above regarding sample selection and the metallicity of any possible \textit{in-situ} halo contamination, the hypothesis, that the GS/E DF has constant anisotropy and the inner-Galaxy decrease in anisotropy is driven by low-anisotropy halo stars, is clearly disfavoured.

\subsection{Comparison with IllustrisTNG50}
\label{subsec:comparison-with-illustris}

We now turn to simulations to further understand whether the trend towards isotropy in the inner Galaxy in the GS/E population is expected in a cosmological context. For this, we follow the approach of \citet{lane25} and use the high resolution TNG50 simulations \citep{tng50_nelson19,tng50_pillepich19}, a component of the IllustrisTNG project \citep{tng_public_release_nelson19}. These authors identified 30 Milky Way analogs in the TNG50 simulation volume, and stellar remnants of 116 major mergers (defined as secondary to primary stellar mass ratio greater than 1:20 at the merger epoch) in the halos of those analogs. Here, we analyze the same remnants to determine whether or not the anisotropy profile of GS/E is typical. We compute the anisotropy profiles of the simulated remnants in a manner similar to the way in which the anisotropy of GS/E is computed. Remnant star particles from the $z=0$ snapshot are selected, and the anisotropy is computed using variable width bins with a constant number of particles per bin. The method is the same as was used by \citet{lane25}, and the interested reader may consult that work for further explanation as well as additional details regarding the selection and preparation of the remnants for study.

\begin{figure}
    \centering
    \includegraphics[width=0.6\columnwidth]{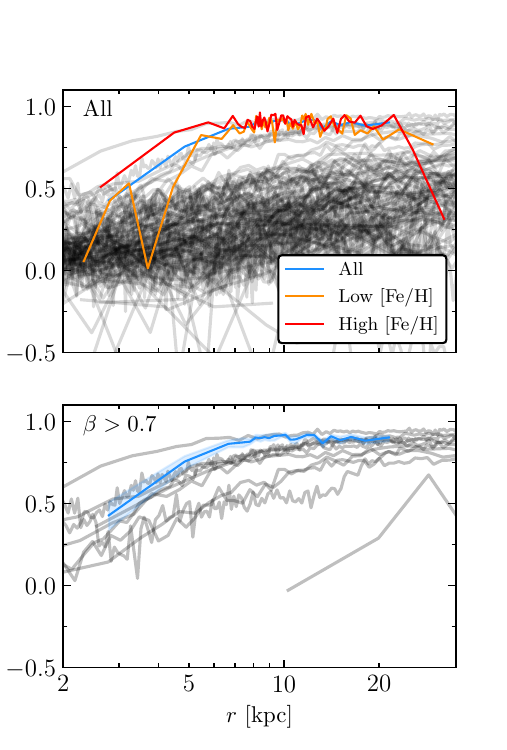}
    \caption{The anisotropy profiles of 116 major merger remnants in 30 Milky Way analogs from the IllustrisTNG simulations. \textit{top:} The anisotropy profiles of all simulated remnants are shown in black and each of the GS/E anisotropy profiles computed in this work are shown in color. \textit{bottom:} Only the anisotropy profiles of the simulated remnants with average $\beta > 0.7$ across their whole radial extent are shown in Black and the anisotropy profile for the whole GS/E sample, along with uncertainties as a colored fill, is shown in Blue for comparison. }
    \label{fig:tng-comparison}
\end{figure}

In Figure~\ref{fig:tng-comparison}, we display the anisotropy profiles of GS/E from Figure~\ref{fig:anisotropy} along with the anisotropy profiles of the 116 simulated merger remnants, computed in a similar manner at $z=0$. While it is clear that the majority of remnants are more isotropic than GS/E, there are a small subset, about 10 or so, which have similar anisotropy profiles. In the bottom panel, we show just those remnants with average anisotropy greater than $\beta=0.7$, along with the combined anisotropy profile for GS/E. The broad similarity between these anisotropy profiles suggests that for GS/E to have an Osipkov-Merritt-like anisotropy profile is by no means out of place in the context of simulated analogs. Importantly, there are no remnants that have high anisotropy in their outer extents for which the anisotropy does not begin to drop in the inner few kpc. Therefore it would actually be surprising to see that the GS/E anisotropy profile is high and constant until its inner extent at a few kpc. This point offers another hurdle for the countertheory entertained above, that the observed lowering of the GS/E anisotropy profile is caused by contamination from \textit{in-situ} populations, as otherwise one might expect to see the implied type of anisotropy profile (high and constant all the way to the inner Galaxy) in these simulations.

\subsection{Reassessing the mass of GS/E}
\label{subsec:reassessing-gse-mass}

In a previous work, \citet{lane23} study GS/E by fitting density profiles to selections of halo stars from APOGEE DR16. They build on the work of \citet{lane22} and use DF-informed kinematic selections to generate high-purity samples of GS/E stars. When fitting density profiles to these samples they account for the bias imposed by the kinematic selections using an assumed DF for GS/E. The correction, in essence, compensates for the fact that to choose a high-purity sample one must necessarily consider a small region of phase space, and so the number of GS/E stars in the sample is smaller than the true size of the GS/E population. \citet{lane23} then compute a series of mass estimates by normalizing the amplitude of the best-fitting density profiles---which requires knowledge of the APOGEE selection function---and then integrating them over an appropriate spatial extent.

The results of both \citet{lane22} and \citet{lane23} rely on an \textit{ansatz} DF for GS/E, the properties of which are sourced from previous works. They use a constant anisotropy DF with $\beta = 0.9$ and a density profile based on the results of \citet{mackereth20}. The choice of a constant anisotropy profile was driven by earlier studies of GS/E, which typically found the anisotropy profile to be constant with a value of approximately 0.9 over a wide range of Galactocentric radii \citep{belokurov18,lancaster19,iorio21}. \citet{lane23} discussed the tension associated with assuming the DF for GS/E in the kinematic selection correction, since the overall goal of the work was itself to determine the mass and density profile of GS/E. To address this they redo their analysis iteratively, using their initial best-fitting GS/E density profile as the basis for the kinematic correction, instead of the assumed form from of \citet{mackereth20}. They found that this exercise did not impact their results or overall conclusions about the physical nature of the GS/E remnant. For this reason, we use the form of the GS/E density profile from \citet{mackereth20} as the basis of the DF-based kinematic selection function here as well. We will see that our updated best-fitting is very similar to the one from \citet{lane23}, so this assumed density profile again does not impact our results.

In light of the results of \citet{lane25}, as well as those presented here, it is natural to inquire whether a change in the assumed DF form for the kinematic corrections would impact the results of \citet{lane23}. Since the role of the DF is both to aid in sample selection and to correct for biases resulting from said selection, the answer to that question is clearly complex. \citet{lane23} argued that their results were resilient to changes in the assumed density profile of the DF based on tests they performed, but they did not study changes in the assumed form of the DF. It is certainly likely that an Osipkov-Merritt-type DF, with lower anisotropy in the inner Galaxy, would result in an increase in the derived mass of GS/E. This follows from the fact that the DF informs the fraction of GS/E stars lying inside the high-purity selection region, which in all cases (selections using a number of different integrals of motion were used) selects stars on the most radial orbits. For a constant anisotropy DF a high fraction of stars lie inside these selection regions. In an Osipkov-Merritt-type DF, where the anisotropy is high and constant in the outer Galaxy and decreases in the inner Galaxy, there is an overall lower fraction of stars on more radial orbits compared with a constant anisotropy DF. 

To test the impact of changing the DF, we re-derive of the stellar mass and density profile of GS/E presented by \citet{lane23} using the best-fitting DF from this work. We are most interested in the stellar mass because it is the property of GS/E which has garnered the most interest over the last few years, with wide range of estimates still accepted \citep[e.g., see ][ for an overview]{deason24}. \citet{lane23} find an uncharacteristically low stellar mass of $\mathrm{M} \sim 1.5 \times 10^{8}~\mathrm{M}_{\odot}$ for GS/E, and so determining if it is actually larger in size would relieve tension between that work and others. The density profile is also of interest, both for studying GS/E in a cosmological context as well as better understanding the structural constituents of the Milky Way.

We repeat the analysis exactly as presented by \citet{lane23}, largely contained in section~3 of that work, except for changes relating to the \textit{ansatz} DF that we outline here. We begin in section~3.2 of \citet{lane23} where they present the kinematic effective selection function. Their equation~(7) shows the \textit{ansatz} constant anisotropy DF with which they construct the kinematic effective selection function \citep[and which is the same DF used by ][]{lane22}. In our analysis we replace this DF with the best-fitting superposition Osipkov-Merritt DF for the whole sample (``All'') presented in Table~\ref{tab:om-fitting-results}. We otherwise keep the analysis the same as presented by \citet{lane23}, including the choice of APOGEE~DR16 data (section 2), the density modelling framework (section~3.1), the form of the kinematic effective selection function (section~3.2, specifically equation~6), the stellar number density models (section~3.3), the fitting procedure (section~3.4), and the computation of the mass (section~3.5). For a full description of the analysis these interested reader is referred to these sections in \citet{lane23}.

The results of our new density profile fits and mass calculations are shown in Table~\ref{tab:structural} \citep[c.f. table~1 in ][]{lane23}. Following \citet{lane23} we fit three different samples defined by kinematic selections in the eccentricity-angular momentum ($e-L_\mathrm{z}$), action diamond (AD), and radial action-angular momentum spaces ($\sqrt{J_\mathrm{r}}-L_\mathrm{z}$). More information on these kinematic selections can be found in section~2.2 of \citet{lane23} and section~3.3 of \citet{lane22}, and the definitions for the kinematic selection rules can be found in table~2 of \citet{lane22}. Eight density profiles are considered, 4 profiles with and without a disk contamination model. The notation used in Tables~\ref{tab:structural} and \ref{tab:likelihood} is as follows: ``SPL'' is a single power law model, ``SC'' is a single power law model with exponential truncation, ``BPL'' is a singly broken power law, ``DBPL'' is a doubly broken power law, and finally ``+D'' refers to a model with an added disk contamination component. The definitions of these density profiles can be found in section~3.3 of \citet{lane23}. Table~\ref{tab:structural} shows the results of fitting each of the density profiles to each of the stellar samples.

In order to gauge the fits to the samples we also follow \citet{lane23} and compute the maximum likelihood $\max(\mathcal{L})$, the Bayesian information criterion $\mathrm{BIC} = k \ln(n) - 2\max(\mathcal{L})$ \citep{schwarz78}, and Akaike's information criterion $\mathrm{AIC} = 2k - 2\max(\mathcal{L})$ \citep{akaike74}. Each of these values for each model and each sample is reported in Table~\ref{tab:likelihood}. We normalize the value for each parameter by the value derived for the single power law model (SPL) for each sample. Since the samples defined by each kinematic selection are different, model fits cannot be compared between kinematic selections. 

We choose a best-fitting model for each kinematic sample with guidance from the information presented in Table~\ref{tab:likelihood}. Higher values of $\max(\mathcal{L})$, and more negative values of AIC and BIC indicate better fits. AIC, and especially BIC, are more reliable parameters for assessing goodness of fit since they appropriately penalize models with larger numbers of parameters. For both the AD and $\sqrt{J_\mathrm{r}}-L_\mathrm{z}$ selections the best model has the lowest AIC and BIC, these being the SC and SPL+D models respectively. For the $e-L_\mathrm{z}$ sample the pattern of AIC and BIC values is the same as was found by \citet{lane23}, namely BIC is better for the SC model but AIC is better for the SC+D model. We therefore follow \citet{lane23} and choose the SC+D model for the $e-L_\mathrm{z}$ sample, noting that the parameters of this model and derived mass are similar to the SC model. We highlight these best-fitting models for each sample in light Gray in both in Tables~\ref{tab:structural} and \ref{tab:likelihood}. 

\begin{figure}
    \centering
    \includegraphics[width=0.6\columnwidth]{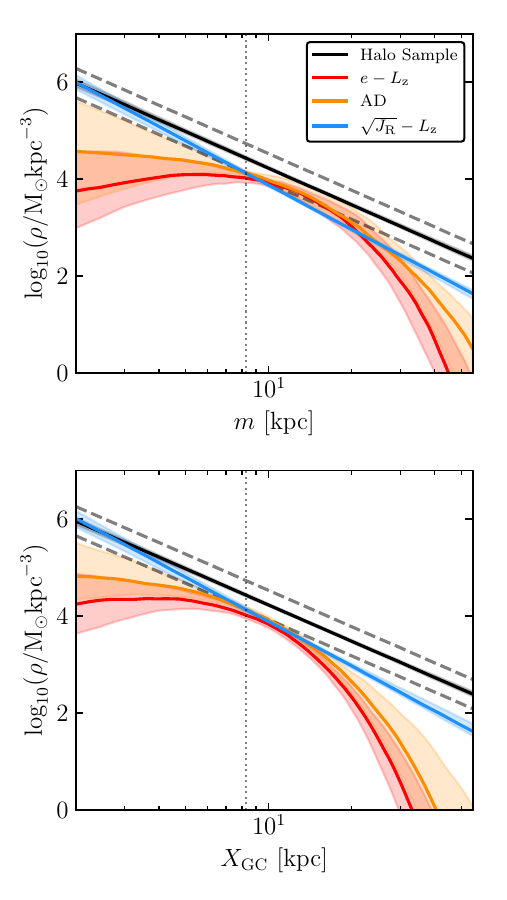}
    \caption{The best-fitting mass density of GS/E based on the reassessed fitting done in this work. This Figure is laid out exactly the same as figure~9 of \citet{lane23}. \textit{top:} mass density as a function of triaxial radius $m$. The red, orange, and blue lines show the three different fits based on the different kinematic selections. Parameters for the density profiles are taken from Table~\ref{tab:structural}, and the density profile forms can be found in section~3.3 of \citet{lane23}. The solid lines show the median density profile, and the surrounding fill the central 68th percentile of 100 samples drawn from the posterior. The black line shows the density profile of the full Halo sample, which is the exact same as shown \textit{lane23} since we do not study the Halo sample in this work. The light gray dashed lines show density profiles twice/half as heavy as the Halo sample profile. \textit{bottom:} mass density as a function of the coordinate proceeding from the Galactic center through the Sun. The colored lines are the same as the top panel.}
    \label{fig:gse-density}
\end{figure}

The best-fitting mass density profiles are shown in Figure~\ref{fig:gse-density}, which is laid out identically to figure~9 in \citet{lane23}. The top panel of this Figure shows the density as a function of $m$, the triaxial radius, and the bottom panel shows it as a function of $X_\mathrm{GC}$, the coordinate along the Galactic center-Sun line. The density profiles are very similar in shape to those presented in figure~9 of \citet{lane23}, but notably the $e-L_\mathrm{z}$ and AD sample models have steeper, less cored central densities. The also have larger exponential cutoff radii, but since their power laws are steeper the appearance is for the profile to be truncated at smaller radii, between 10--20~kpc. The best-fitting SPL model for the $\sqrt{J_\mathrm{r}}-L_\mathrm{z}$ sample is steeper than was found by \citet{lane23}. These small differences aside, interpretation of these density profiles is not significantly different than the interpretation offered by \citet{lane23} (see their section~5.2.3). To briefly summarize: the $e-L_\mathrm{z}$ and AD sample models show a flat inner density profile which is effectively truncated around 20~kpc, consistent with previous findings \citep{deason18,han22}. \citet{lane23} argues, as we do here as well, that the similarity between the best-fitting $\sqrt{J_\mathrm{r}}-L_\mathrm{z}$ SPL model and the whole Halo profile (c.f. the blue and black lines in Figure~\ref{fig:gse-density}) indicates that the $\sqrt{J_\mathrm{r}}-L_\mathrm{z}$ sample suffers from higher contamination from non-GS/E stars than the other samples, a conclusion supported by the results of \citet{lane22}.

\citet{lane23} choose the mass of the best-fitting AD profile to report as the mass for GS/E (see their section~4.3 and 4.5), and we do the same here. We therefore find the mass of GS/E to be $M = 2.29^{+0.95}_{-0.63} \times 10^{8}~\mathrm{M}_{\odot}$. This value is about 50~per~cent larger than the value reported by \citet{lane23} for the AD sample. For the best-fitting $e-L_\mathrm{z}$ sample profile the mass found here is about 40~per~cent larger than that found by \citet{lane23}. Interestingly the mass of the best-fitting $\sqrt{J_\mathrm{r}}-L_\mathrm{z}$ sample profile is about 15~per~cent smaller than that found by \citet{lane23}, which is likely due to the steeper density profile found here which will carry much lower mass at large radii. Since the $e-L_\mathrm{z}$ and AD samples are favoured by \citet{lane23} on grounds of purity, their agreement in the elevation of the GS/E stellar mass is promising. The overall trend of masses to increase when using the new superposition Osipkov-Merritt DF for the kinematic effective selection function is consistent with the expectations outlined previously in this section.

This result is interesting because the mass of GS/E reported by \citet{lane23} is quite low compared with any other estimate in the literature. The new mass of $\sim 2.3 \times 10^{8}~\mathrm{M}_{\odot}$ is closer to many other estimates on the low end of the spectrum \citep[e.g.,][]{mackereth20,rey23,carrillo24}, especially those derived using globular clusters \citep{kruijssen19,callingham22}. This agreement helps to reinforce the idea that GS/E probably constituted a less impactful merger in the history of the Milky Way than has been traditionally thought, however the mass is not likely to be as low as originally derived by \citet{lane23}.

\section{Conclusion}
\label{sec:conclusion}

GS/E is one of the most important remnants in the Milky Way stellar halo, and has received much attention over the last few years. Recent work by \citet{lane25} found that simulated accretion remnants with high average anisotropy tend to have anisotropy profiles which resemble that sourced from an Osipkov-Merritt-type DF. Here, we study the anisotropy profile of GS/E using a sample of chemically-selected stars from APOGEE DR17 and \textit{Gaia} DR3. We use a chemistry-based selection protocol to avoid introducing kinematic biases that may cloud interpretation of the GS/E anisotropy profile. Our selection begins with a locus in [Fe/H] and [Mg/Fe] defined similarly to other selections in the literature. We also add a cut in [Al/Fe] to remove \textit{in-situ} stars, and finally add a very restrictive cut in [Fe/H] and $v_\mathrm{circ}$ to excise a concentrated group of disk stars.

We analyze the anisotropy profile of the resulting sample, finding that it exhibits the characteristics of an Osipkov-Merritt anistoropy profile. It is elevated to $\beta=0.9$ at $r > 8$~kpc, consistent with previous assessments of the anisotropy profile that focus on the solar circle and outer Galaxy. But within 8~kpc the anisotropy profile drops smoothly to $\beta=0.4$. As a simple approximate representation  of the underlying distribution function of GS/E, we fit a superposition Osipkov-Merritt anisotropy profile to the data, finding it is able to reproduce the anisotropy profile at large radii, and also do a reasonable job of following the decreasing anisotropy at small radii. Of course, as a spherical model, while these Osipkov-Merritt-type models can reproduce the radial density and anisotropy profiles, they cannot reproduce the triaxial shape and kinematics of GS/E, which is a limitation to keep in mind when using this result.

We test the resilience of our findings to choices in selection protocol, and find that using both less restrictive and more restrictive chemical and kinematic cuts still leads to a sample with an anisotropy profile with Osipkov-Merritt-like characteristics. We also investigate a possible alternative hypothesis that could explain our observations, namely that the anisotropy decrease in the inner Galaxy is driven by \textit{in-situ} halo contamination, and that GS/E actually has a constant anisotropy profile to as small radii as we are able to probe. We test this hypothesis by creating and sampling from a candidate DF with the correct anisotropy profile based on the \textit{Aurora} halo population combined with a constant anisotropy DF for GS/E. We determine that this DF does not satisfactorily match the GS/E data when examined using other kinematic properties, and that our best-fitting Osipkov-Merritt superposition DF does a better job, allowing us to refute the contamination hypothesis.

We summarize our findings here:

\begin{itemize}
    \item We produce a sample of 2,367 likely GS/E stars from APOGEE DR17 using a chemical selection based on [Fe/H], [Mg/Fe], and [Al/Fe] abundances.
    \item The anisotropy profile of our GS/E sample is elevated at $r > 8$~kpc, and drops to $\beta \sim 0.4$ moving towards 2~kpc, the innermost radii probed by our sample. Such an anisotropy profile is consistent with an Osipkov-Merritt-type DF.
    \item As a simple approximate model, we fit a superposition Osipkov-Merritt anisotropy profile to the GS/E data. We find that a two-component model with scale radii $r_{\mathrm{a},1}=2$~kpc, $r_{\mathrm{a},2} = 547$~kpc, and a mixture fraction of $k_\mathrm{om}=0.88$ provides the best fit.
    \item We demonstrate that our results survive modification to the selection protocol used to create the GS/E sample. We also investigate whether an alternative hypothesis, that the observed decrease in the anisotropy profile is caused by \textit{in-situ} halo contamination, could better explain our observations, and find that it cannot.
    \item We compare the anisotropy profiles of simulated accretion remnants around Milky Way analogs from the IllustrisTNG simulations with the GS/E data. Simulated remnants that are comparable to GS/E (i.e., they have high constant anisotropy of about $\beta=0.9$) also tend to exhibit drops in their anisotropy near the center of the simulated host galaxy. This drop occurs at similar radii and is of a similar magnitude to what is observed in the GS/E data, suggesting that the anisotropy profile of GS/E fits in well among its simulated peers. 
    \item We re-assess the stellar mass and density profile of the GS/E remnant found by \citet{lane23}, who relied on an assumed constant anisotropy for GS/E to aid in their analysis. We repeat their analysis, but change the DF to used in the kinematic effective selection function to be the best-fitting superposition Osipkov-Merritt DF from this work. We re-fit density profiles to several samples defined by different kinematic selections, arriving at very similar results as \citet{lane23}. Computing the mass for the best-fitting density profile yields a value of $2.29^{+0.95}_{-0.63}\times 10^{8}~\mathrm{M}_{\odot}$ for the stellar mass of GS/E. This value is a factor of about 50~per~cent larger than the value of $1.5 \times 10^{8}~\mathrm{M}_{\odot}$ originally presented by \citet{lane23}, and it puts the estimate in better agreement with other low-end stellar mass estimates for GS/E.
\end{itemize}

This work has revealed key information about the nature of GS/E, perhaps the most important stellar population in the Milky Way halo. Our rederivation of the mass of GS/E, which is now firmly placed among other low-end estimates for the mass, will help to reinforce the notion that GS/E likely had a slightly smaller mass progenitor than is typically thought. Furthermore, future modelling of the remnant can now use a superposition Osipkov-Merritt model to represent GS/E, which will provide a more accurate basis to perform work such as that of \citet{lane23} that relies on knowledge of the DF. Finally, our investigation of the kinematics of GS/E in the inner Milky Way will help to place important constraints on the next generation of efforts to model the accretion event which produced GS/E as well as the historical evolution of the remnant.

\begin{acknowledgments}

We thank the anonymous referee for a thoughtful report. JMML and JB acknowledge financial support from NSERC (funding reference number RGPIN-2020-04712).

\end{acknowledgments}

\appendix

\section{Results of new density profile fits to GS/E}

Here we present the results of the refits to the APOGEE DR16 data following \citet{lane23} and presented in \S~\ref{subsec:reassessing-gse-mass}. Table~\ref{tab:structural} shows the density profile fits to GS/E for the three different kinematic selections. This table is identical in form to table~1 of \citet{lane23}, except that the results for the entire halo are not included, since its analysis is not changed by the introduction of a new DF for GS/E. Rows highlighted in light-Gray indicate the best-fitting density profile for that sample. Table~\ref{tab:likelihood} shows the maximum likelihood, Bayesian Information Criterion (BIC), and Akaike Information Criterion (AIC) for each of the models under consideration. This table is identical in form to table~2 of \citet{lane23}. The interested reader should consult \citet{lane23} for additional information on how these results are obtained.

\begin{deluxetable}{lcccccccccccc}
\tabletypesize{\scriptsize}
\tablecolumns{13}
\tablehead{\colhead{Name} & \colhead{$\alpha_{1}$} & \colhead{$\alpha_{2}$} & \colhead{$\alpha_{3}$} & \colhead{$r_{1}$} & \colhead{$r_{2}$} & \colhead{$p$} & \colhead{$q$} & \colhead{$\theta$} & \colhead{$\eta$} & \colhead{$\phi$} & \colhead{$f_\mathrm{disk}$} & \colhead{Mass} \\ \colhead{} & \colhead{} & \colhead{} & \colhead{} & \colhead{$[\mathrm{kpc}]$} & \colhead{$[\mathrm{kpc}]$} & \colhead{} & \colhead{} & \colhead{$[\mathrm{deg}]$} & \colhead{} & \colhead{$[\mathrm{deg}]$} & \colhead{} & \colhead{$[10^{8}~M_{\odot}]$}}
\tablecaption{Results of the fits to the APOGEE DR16 data following the approach of \citet{lane23} (c.f. their table~1) but using the new Osipkov-Merritt DF based on the results in this work.\label{tab:structural}}
\startdata
\cutinhead{$e-L_\mathrm{z}$}
SPL & $3.47^{+0.21}_{-0.22}$ & $-$ & $-$ & $-$ & $-$ & $0.93^{+0.05}_{-0.08}$ & $0.57^{+0.08}_{-0.07}$ & $112.8^{+188.1}_{-58.7}$ & $0.99^{+0.01}_{-0.02}$ & $89.9^{+66.3}_{-66.1}$ & $-$ & $8.33^{+0.04}_{-0.04}$ \\ 
SPL+D & $3.28^{+0.25}_{-0.26}$ & $-$ & $-$ & $-$ & $-$ & $0.91^{+0.06}_{-0.1}$ & $0.72^{+0.16}_{-0.14}$ & $132.3^{+175.5}_{-82.0}$ & $0.95^{+0.04}_{-0.22}$ & $86.1^{+62.8}_{-56.5}$ & $0.66^{+0.12}_{-0.22}$ & $8.3^{+0.05}_{-0.05}$ \\ 
SC & $-0.55^{+1.29}_{-1.3}$ & $-$ & $-$ & $3.37^{+1.42}_{-0.78}$ & $-$ & $0.8^{+0.11}_{-0.1}$ & $0.52^{+0.09}_{-0.08}$ & $81.1^{+206.8}_{-47.9}$ & $0.99^{+0.01}_{-0.02}$ & $96.6^{+15.3}_{-16.0}$ & $-$ & $8.09^{+0.06}_{-0.06}$ \\ 
$^*$SC+D & $-1.76^{+1.59}_{-1.68}$ & $-$ & $-$ & $2.98^{+1.19}_{-0.66}$ & $-$ & $0.71^{+0.13}_{-0.11}$ & $0.54^{+0.13}_{-0.1}$ & $81.1^{+169.6}_{-56.5}$ & $0.97^{+0.02}_{-0.1}$ & $96.3^{+12.0}_{-13.3}$ & $0.64^{+0.12}_{-0.21}$ & $8.03^{+0.07}_{-0.06}$ \\ 
BPL & $1.83^{+0.59}_{-0.79}$ & $5.37^{+1.1}_{-0.84}$ & $-$ & $13.28^{+2.78}_{-2.51}$ & $-$ & $0.81^{+0.12}_{-0.11}$ & $0.52^{+0.09}_{-0.08}$ & $93.3^{+227.1}_{-63.1}$ & $0.99^{+0.01}_{-0.02}$ & $98.0^{+17.0}_{-19.5}$ & $-$ & $8.11^{+0.07}_{-0.06}$ \\ 
BPL+D & $1.29^{+0.73}_{-0.79}$ & $5.25^{+1.2}_{-0.88}$ & $-$ & $13.8^{+3.6}_{-2.82}$ & $-$ & $0.75^{+0.13}_{-0.13}$ & $0.56^{+0.15}_{-0.11}$ & $91.2^{+219.9}_{-66.4}$ & $0.98^{+0.02}_{-0.11}$ & $99.1^{+16.6}_{-16.8}$ & $0.62^{+0.13}_{-0.23}$ & $8.06^{+0.08}_{-0.08}$ \\ 
DBPL & $1.64^{+0.66}_{-0.85}$ & $4.81^{+1.1}_{-1.17}$ & $7.21^{+1.89}_{-1.68}$ & $12.39^{+2.78}_{-2.89}$ & $29.61^{+16.9}_{-12.62}$ & $0.81^{+0.11}_{-0.11}$ & $0.52^{+0.09}_{-0.08}$ & $84.8^{+230.8}_{-55.5}$ & $0.99^{+0.01}_{-0.02}$ & $98.4^{+16.8}_{-17.9}$ & $-$ & $8.1^{+0.06}_{-0.06}$ \\ 
DBPL+D & $1.04^{+0.85}_{-0.72}$ & $4.52^{+1.13}_{-1.38}$ & $7.17^{+1.87}_{-1.73}$ & $12.6^{+3.35}_{-2.81}$ & $28.19^{+16.64}_{-10.5}$ & $0.74^{+0.13}_{-0.12}$ & $0.57^{+0.15}_{-0.11}$ & $98.0^{+208.0}_{-72.1}$ & $0.98^{+0.02}_{-0.1}$ & $99.3^{+15.2}_{-16.6}$ & $0.61^{+0.14}_{-0.24}$ & $8.05^{+0.07}_{-0.07}$ \\ 
\cutinhead{AD}\\ 
SPL & $3.27^{+0.3}_{-0.3}$ & $-$ & $-$ & $-$ & $-$ & $0.77^{+0.14}_{-0.14}$ & $0.69^{+0.19}_{-0.2}$ & $135.5^{+61.4}_{-77.8}$ & $0.83^{+0.12}_{-0.21}$ & $101.7^{+21.8}_{-22.6}$ & $-$ & $8.61^{+0.1}_{-0.09}$ \\ 
SPL+D & $3.14^{+0.33}_{-0.33}$ & $-$ & $-$ & $-$ & $-$ & $0.71^{+0.17}_{-0.17}$ & $0.65^{+0.21}_{-0.2}$ & $142.5^{+50.7}_{-92.1}$ & $0.79^{+0.14}_{-0.19}$ & $99.8^{+19.9}_{-19.3}$ & $0.51^{+0.21}_{-0.29}$ & $8.6^{+0.1}_{-0.1}$ \\ 
$^*$SC & $0.13^{+1.92}_{-2.03}$ & $-$ & $-$ & $5.66^{+8.8}_{-2.34}$ & $-$ & $0.66^{+0.16}_{-0.14}$ & $0.58^{+0.22}_{-0.18}$ & $144.1^{+41.6}_{-89.8}$ & $0.84^{+0.11}_{-0.21}$ & $100.4^{+13.9}_{-14.7}$ & $-$ & $8.36^{+0.15}_{-0.14}$ \\ 
SC+D & $0.31^{+1.79}_{-1.97}$ & $-$ & $-$ & $6.66^{+11.64}_{-3.0}$ & $-$ & $0.63^{+0.17}_{-0.16}$ & $0.56^{+0.22}_{-0.18}$ & $147.2^{+38.3}_{-99.2}$ & $0.81^{+0.13}_{-0.2}$ & $99.9^{+14.1}_{-14.9}$ & $0.43^{+0.23}_{-0.27}$ & $8.36^{+0.16}_{-0.14}$ \\ 
BPL & $1.08^{+1.34}_{-0.77}$ & $4.49^{+1.07}_{-0.66}$ & $-$ & $13.85^{+7.29}_{-2.68}$ & $-$ & $0.69^{+0.15}_{-0.14}$ & $0.61^{+0.2}_{-0.18}$ & $149.1^{+44.8}_{-97.2}$ & $0.83^{+0.12}_{-0.22}$ & $100.6^{+14.9}_{-15.9}$ & $-$ & $8.36^{+0.13}_{-0.12}$ \\ 
BPL+D & $1.32^{+1.37}_{-0.94}$ & $4.44^{+1.53}_{-0.75}$ & $-$ & $15.42^{+12.25}_{-3.8}$ & $-$ & $0.65^{+0.17}_{-0.16}$ & $0.59^{+0.21}_{-0.18}$ & $149.3^{+40.9}_{-106.3}$ & $0.8^{+0.14}_{-0.2}$ & $99.7^{+14.8}_{-15.5}$ & $0.41^{+0.24}_{-0.25}$ & $8.38^{+0.14}_{-0.13}$ \\ 
DBPL & $0.94^{+1.04}_{-0.67}$ & $4.05^{+0.82}_{-0.85}$ & $6.76^{+2.16}_{-1.95}$ & $12.89^{+3.89}_{-2.65}$ & $34.45^{+13.5}_{-12.3}$ & $0.68^{+0.15}_{-0.14}$ & $0.61^{+0.2}_{-0.18}$ & $150.7^{+44.8}_{-99.3}$ & $0.83^{+0.12}_{-0.22}$ & $100.0^{+14.4}_{-15.4}$ & $-$ & $8.33^{+0.12}_{-0.12}$ \\ 
DBPL+D & $1.06^{+1.1}_{-0.76}$ & $3.94^{+0.87}_{-0.85}$ & $6.79^{+2.21}_{-2.06}$ & $13.4^{+5.62}_{-2.93}$ & $35.33^{+13.0}_{-11.68}$ & $0.65^{+0.17}_{-0.15}$ & $0.6^{+0.22}_{-0.18}$ & $149.1^{+43.3}_{-109.4}$ & $0.8^{+0.14}_{-0.2}$ & $99.4^{+15.2}_{-15.6}$ & $0.4^{+0.23}_{-0.25}$ & $8.34^{+0.13}_{-0.12}$ \\ 
\cutinhead{$\sqrt{J_\mathrm{R}}-L_\mathrm{z}$} \\ 
SPL & $3.17^{+0.17}_{-0.17}$ & $-$ & $-$ & $-$ & $-$ & $0.7^{+0.07}_{-0.07}$ & $0.7^{+0.08}_{-0.07}$ & $170.2^{+46.2}_{-150.6}$ & $0.84^{+0.13}_{-0.22}$ & $106.5^{+8.1}_{-8.1}$ & $-$ & $8.72^{+0.03}_{-0.03}$ \\ 
$^*$SPL+D & $3.03^{+0.18}_{-0.19}$ & $-$ & $-$ & $-$ & $-$ & $0.66^{+0.08}_{-0.07}$ & $0.69^{+0.1}_{-0.08}$ & $171.0^{+61.7}_{-146.7}$ & $0.9^{+0.08}_{-0.23}$ & $105.6^{+7.4}_{-7.5}$ & $0.6^{+0.11}_{-0.18}$ & $8.71^{+0.04}_{-0.04}$ \\ 
SC & $2.66^{+0.26}_{-0.36}$ & $-$ & $-$ & $34.04^{+14.26}_{-14.19}$ & $-$ & $0.69^{+0.07}_{-0.07}$ & $0.7^{+0.08}_{-0.07}$ & $167.9^{+46.4}_{-148.3}$ & $0.85^{+0.12}_{-0.22}$ & $106.9^{+7.6}_{-7.7}$ & $-$ & $8.64^{+0.04}_{-0.05}$ \\ 
SC+D & $2.43^{+0.3}_{-0.44}$ & $-$ & $-$ & $32.42^{+15.04}_{-14.11}$ & $-$ & $0.64^{+0.08}_{-0.07}$ & $0.69^{+0.11}_{-0.09}$ & $169.1^{+69.0}_{-141.4}$ & $0.92^{+0.06}_{-0.22}$ & $106.2^{+6.8}_{-7.1}$ & $0.62^{+0.11}_{-0.17}$ & $8.61^{+0.05}_{-0.06}$ \\ 
BPL & $2.52^{+0.69}_{-1.5}$ & $3.52^{+2.38}_{-0.3}$ & $-$ & $12.31^{+34.34}_{-4.86}$ & $-$ & $0.69^{+0.07}_{-0.07}$ & $0.7^{+0.09}_{-0.07}$ & $169.5^{+45.3}_{-149.7}$ & $0.84^{+0.13}_{-0.22}$ & $106.8^{+7.9}_{-8.1}$ & $-$ & $8.65^{+0.06}_{-0.07}$ \\ 
BPL+D & $1.75^{+1.21}_{-1.2}$ & $3.32^{+0.79}_{-0.28}$ & $-$ & $11.52^{+27.68}_{-3.38}$ & $-$ & $0.63^{+0.08}_{-0.07}$ & $0.69^{+0.11}_{-0.09}$ & $175.2^{+66.5}_{-146.7}$ & $0.91^{+0.07}_{-0.21}$ & $106.9^{+7.0}_{-7.4}$ & $0.61^{+0.11}_{-0.16}$ & $8.61^{+0.07}_{-0.07}$ \\ 
DBPL & $1.78^{+1.22}_{-1.14}$ & $3.3^{+0.4}_{-0.61}$ & $4.56^{+3.09}_{-1.12}$ & $9.21^{+10.43}_{-4.52}$ & $41.12^{+10.22}_{-28.59}$ & $0.69^{+0.07}_{-0.07}$ & $0.7^{+0.09}_{-0.08}$ & $180.3^{+41.8}_{-158.1}$ & $0.85^{+0.12}_{-0.22}$ & $107.5^{+7.6}_{-7.8}$ & $-$ & $8.61^{+0.07}_{-0.07}$ \\ 
DBPL+D & $1.42^{+1.26}_{-0.95}$ & $3.2^{+0.34}_{-0.33}$ & $5.1^{+3.11}_{-1.64}$ & $10.03^{+5.84}_{-3.66}$ & $44.83^{+7.42}_{-20.16}$ & $0.64^{+0.08}_{-0.07}$ & $0.71^{+0.11}_{-0.09}$ & $176.6^{+68.8}_{-140.1}$ & $0.92^{+0.06}_{-0.2}$ & $107.3^{+7.0}_{-7.3}$ & $0.61^{+0.11}_{-0.16}$ & $8.57^{+0.08}_{-0.07}$ \\ 
\enddata
\tablecomments{The names refer to density profiles as follows: `SPL' single power law; `SC' exponentially truncated single power law; `BPL' broken power law; `DBPL' double broken power law; `+D' refers to an added disk contamination component. The table is divided into three sections, one for each kinematic selection used to study GS/E. The starred rows indicate the best-fitting density profile for each sample.}
\end{deluxetable}

\begin{deluxetable*}{lccc}[ht!]
\centerwidetable
\tablecolumns{4}
\tablehead{\colhead{Name} & \colhead{$\mathrm{max}(\mathcal{L})$} & \colhead{AIC} & \colhead{BIC}}
\tablecaption{Maximum likelihoods, BIC, and AIC for each model and each kinematic selection.\label{tab:likelihood}}
\tablewidth{\textwidth}
\startdata
\cutinhead{$e-L_\mathrm{z}$} \\ 
SPL & 0.0 & 0.0 & 0.0 \\ 
SPL+D & 1.0 & $-0.1$ & 3.0 \\ 
SC & 10.8 & $-19.5$ & $-16.5$ \\ 
$^*$SC+D & 13.0 & $-22.1$ & $-16.0$ \\ 
BPL & 10.5 & $-17.1$ & $-11.0$ \\ 
BPL+D & 12.1 & $-18.2$ & $-9.2$ \\ 
DBPL & 11.6 & $-15.3$ & $-3.2$ \\ 
DBPL+D & 14.4 & $-18.7$ & $-3.6$ \\ 
\cutinhead{AD} \\ 
SPL & 0.0 & 0.0 & 0.0 \\ 
SPL+D & 0.9 & 0.1 & 2.4 \\ 
$^*$SC & 6.1 & $-10.2$ & $-7.9$ \\ 
SC+D & 6.2 & $-8.4$ & $-3.8$ \\ 
BPL & 7.0 & $-10.0$ & $-5.4$ \\ 
BPL+D & 7.1 & $-8.2$ & $-1.4$ \\ 
DBPL & 7.2 & $-6.5$ & 2.7 \\ 
DBPL+D & 7.7 & $-5.3$ & 6.1 \\ 
\cutinhead{$\sqrt{J_\mathrm{R}}-L_\mathrm{z}$} \\ 
SPL & 0.0 & 0.0 & 0.0 \\ 
$^*$SPL+D & 4.0 & $-6.0$ & $-2.3$ \\ 
SC & 0.4 & 1.1 & 4.8 \\ 
SC+D & 5.0 & $-6.0$ & 1.4 \\ 
BPL & 2.9 & $-1.8$ & 5.6 \\ 
BPL+D & 7.7 & $-9.4$ & 1.8 \\ 
DBPL & 2.9 & 2.2 & 17.0 \\ 
DBPL+D & 7.6 & $-5.3$ & 13.2 \\ 
\enddata
\tablecomments{This table is identical in layout and construction to table~2 of \citet{lane23} except shows results from the new analysis presented in \S~\ref{subsec:reassessing-gse-mass}. Each value is normalized by the value of the SPL model for the corresponding kinematic selection (which defines the sample to be fit). Larger maximum likelihoods, and more negative BIC and AIC indicate better fits. The starred rows indicate the best-fitting density profile for each sample. Since each kinematic selection defines a different sample, comparison of values between kinematic selection is not advised.}
\end{deluxetable*}

\bibliography{manuscript}{}
\bibliographystyle{aasjournalv7}



\end{document}